\documentclass[aps,prc,twocolumn,amsmath,amssymb,amsfonts,superscriptaddress]{revtex4-1}
\usepackage [final]{graphics}
\usepackage[font=small, justification=raggedright, singlelinecheck=false]{caption}
\usepackage{booktabs,caption}
\usepackage{threeparttable}
\usepackage{ragged2e}
\usepackage {float}
\usepackage {subfig}
\usepackage{epstopdf}
\usepackage{epsfig}
\usepackage{csquotes}
\usepackage{array}
\usepackage{booktabs}
\usepackage{ulem}
\usepackage[hang]{footmisc}
\begin{document}

\title{Stellar $s$--process neutron capture cross section of Ce isotopes}
\author{R. N. Sahoo}
\affiliation{The Hebrew University of Jerusalem, Jerusalem, Israel 91904 }
\author{M. Paul}
\email{paul@vms.huji.ac.il}
\affiliation{The Hebrew University of Jerusalem, Jerusalem, Israel 91904 }
\author{Y. Kashiv}
\affiliation{University of Notre Dame, Notre Dame, IN 46556, United States}
\author{M. Tessler}
%\email{moshe.tessler@mail.huji.ac.il}
\affiliation{Soreq Nuclear Research Center, Yavne, Israel 81800 }
\affiliation{The Hebrew University of Jerusalem, Jerusalem, Israel 91904 }
\author{\\M. Friedman}
\affiliation{The Hebrew University of Jerusalem, Jerusalem, Israel 91904 }
\author{S. Halfon}
\affiliation{Soreq Nuclear Research Center, Yavne, Israel 81800 }
%\author{Dani Kijel}
%\affiliation{Soreq Nuclear Research Center, Yavne, Israel 81800 }
\author{A. Kreisel}
\affiliation{Soreq Nuclear Research Center, Yavne, Israel 81800 }
\author{A. Shor}
\affiliation{Soreq Nuclear Research Center, Yavne, Israel 81800 }
\author{L. Weissman}
\affiliation{Soreq Nuclear Research Center, Yavne, Israel 81800 }
%\date{\today}

\begin{abstract}
Stellar abundances of cerium are of high current interest based both on observations and theoretical models, especially with regard to the neutron--magic $^{140}$Ce isotope. A large discrepancy of $s-$process stellar models relative to cerium abundance observed in globular clusters was highlighted, pointing to possible uncertainties in experimental nuclear reaction rates. In 
 this work, 
 the stellar neutron capture cross section of the stable cerium isotopes $^{136}$Ce, $^{138}$Ce, $^{140}$Ce, and $^{142}$Ce, 
 were re-measured. 
 A $^{nat}$Ce sample was irradiated with quasi-Maxwellian neutrons at $kT = 34.2$ keV using the $^{7}$Li($p,n$) reaction. The neutron field with an intensity of $3-5 \times 10^{10}$ n/s was produced by irradiating the liquid-lithium target (LiLiT) with a mA proton beam at an energy (1.92 MeV) just above the threshold at Soreq Applied Research Accelerator Facility (SARAF). The activities of the $^{nat}$Ce 
 neutron capture products
 were measured using a shielded  High Purity Germanium  detector. 
 %Correction was made for ($\gamma,n$) reaction products, inteferring in our setup %with $(n,\gamma)$ production.
 Cross sections were extracted relative to that of the $^{197}$Au(n,$\gamma$) reaction and  
 the 
 Maxwellian-averaged cross section (MACS) of the Ce isotopes were derived.  
 The 
 MACS values 
 extracted from this experiment are generally consistent with previous measurements and show for $^{140}$Ce a value $\approx 15$\% smaller than most recent experimental values.  
 
\end{abstract}
%\pacs{}
\maketitle

\section{\label{sec:level0}Introduction}
\begin{figure}[h!]
\centering
\includegraphics[width=8cm, trim={0.0cm 0.0cm 0.0cm 0cm}]{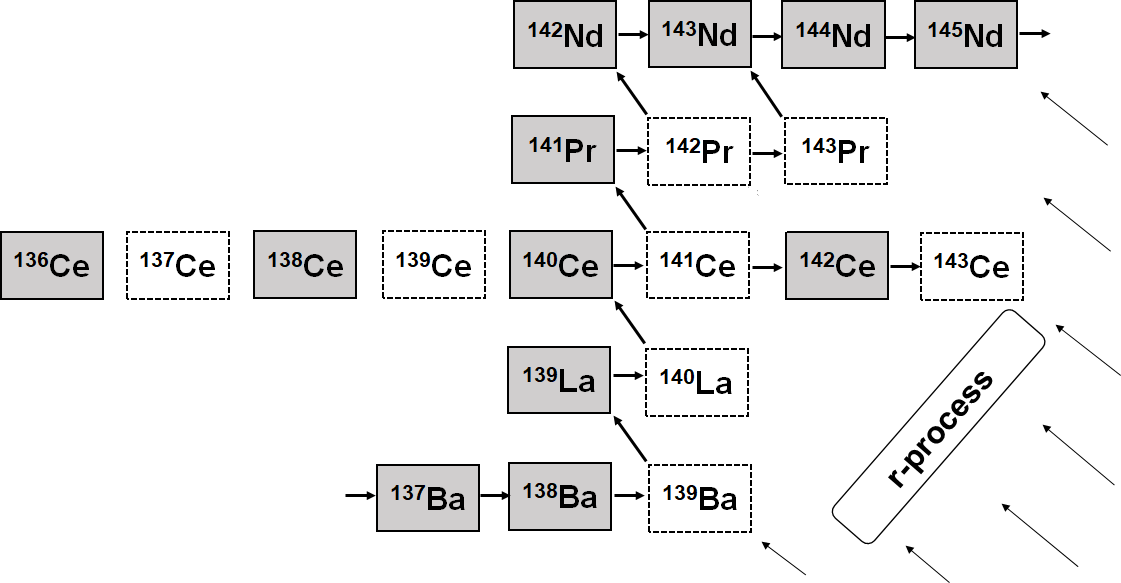}
\caption{Part of the nuclide chart showing the s-process nucleosynthesis path in the Ce-Pr-Nd mass around the branching at A$=141-142$.}
\label{fig:Path_s_process}
\end{figure}

%%%%%%%%%%%%%%%%%%%%%%%%%%%%%%%%%%%%%%%%%%%%%%%%%%%%
The four stable isotopes of cerium (Z = 58, Fig. \ref{fig:Path_s_process}) display a showcase of processes responsible for production of heavy elements. Since Cameron \cite{CAM57} and Burbidge et al. \cite{B2FH}, neutron captures are considered the main production gateways for these nuclides via the slow 
($s$) and the rapid ($r$) processes 
while very few proton-rich nuclides (two of those being Ce isotopes) belong to the so-called 
$p$ process. 
%%%%%%%%%%%%%%%%%%%%%%%%%%%%%%%%%%%%%%%%%%%%%%%%%%%%
 About half of the isotopes of the heavy elements ($A\gtrapprox 60$) are produced by the slow ($s$) process. There, the time between consecutive neutron captures is long with respect to typical $\beta ^-$ decay lifetimes along its path, resulting in an evolution close to the valley of stability. The $s$ process is further divided in a  weak and main components. The weak 
 $s$ process \cite{Kappeler11} 
 operates in massive stars, $M_{\star}>8M_{\odot}$, before they explode as type II supernovae and 
 %fully or partially 
 produces nuclides in the mass range $60\lessapprox A \lessapprox 90$. 
 %and will not be discussed further in this paper.
 The main component \cite{Kappeler11,Lugaro03}, which produces nuclides in the range $90\lessapprox A\leq209$, takes place in low-mass asymptotic giant branch (AGB) stars, $M_\star\leq4M_\odot$. It operates on a timescale of a few $10^5$ yr. In a simplified view, most of the  neutron exposure experienced by seed nuclei is generated by the $^{13}{\mathrm C}(\alpha,n)^{16}{\mathrm O}$ neutron source reaction, operating in pulses of a few $10^4$ yr. The reaction is activated at $T\sim0.9\times10^8$ K ($\approx 8$ keV) and generates a low neutron number density of $N_n=10^6-10^8$ cm$^{-3}$. 
 %As a result, the $s$ process progresses along the valley of $\beta$ stability. 
At the end of such a pulse  when the temperature reaches $T\sim2.5\times10^8$ K ($\approx 22$ keV), a much shorter pulse is activated for a few years during  which the $^{22}\mathrm{Ne}(\alpha,n)^{25}\mathrm{Mg}$ is the main neutron source. It generates a higher neutron number density of $N_n\sim10^{10}$ cm$^{-3}$. Due to the short time it operates, the latter neutron source accounts for only a small fraction of the total exposure. However, the higher $N_n$ can affect branching points along the main $s$-process path.
 The other about half of heavy nuclides  are produced by the rapid ($r$) process, in which the time between consecutive neutron captures is short with respect to radioisotope $\beta ^-$ decay lifetimes along its path. The $r$ process operates on a timescale of 1 sec under explosive conditions characterized by high temperature, $T>10^9$\,K $(\approx90$\,keV), and high neutron number density, $N_n>10^{20}$ cm$^{-3}$ \cite{Freiburghaus99}. 
 %As a result, it progresses close to the neutron drip line. The $r$ process fully/partially produces isotopes up to U and heavier actinides. The astronomical site(s) of the $r$ process has been a matter of debate for many years. Several sites have been proposed, including supernovae and neutron star mergers (NSM). The fact that the electromagnetic spectrum of the kilonova following the GW170817 NSM, the first multimessenger event including gravitational waves \cite{Abbott17} (the electromagnetic event was labeled AT 2017gfo), is best fitted with $r$-process elements \cite{Pian17}, indicates that NSM is at least one of the $r$-%process sites \cite{Cowan21}.
A third 
neutron capture process, termed the $i$ process, was proposed by \cite{Cowan77}. The neutron number density in this process, $N_n=10^{12}-10^{16}$ cm$^{-3}$, is intermediate between those the of 
$r$ and $s$ processes. It is considered to operate in some special cases \cite{Cowan77,Denissenkov19,Choplin21,Choplin22b}. Potential confirmation of the $i$ process 
comes from the chemical compositions of carbon-enhanced metal-poor (CEMP)$-r/s$ stars, enriched in both $r-$ and $s-$process elements.
%The neutron source is the $^{13}{\mathrm C}(\alpha,n)^{16}{\mathrm O}$ reaction, which is activated at a high temperature of $T>2\times10^8$ K ($\approx17$ keV). Potential confirmation of the $i$ process comes from the chemical compositions of carbon-enhanced metal-poor, (CEMP)-$r/s$, stars. These stars are enriched in both $r$- and $s$-process elements, and their compositions are best explained by yields of the $i$ process %\cite{Denissenkov19,Choplin21,Karinkuzhi21}.
In addition to the neutron capture processes described above, the $p$ process produces 35 neutron-deficient stable isotopes. It operates in massive stars under explosive conditions by photonuclear reactions $\textit{e.g.}$ $(\gamma,\alpha\setminus{n}\setminus{p})$, on pre-existing $r$ and $s$ isotopes \cite{Meyer94,Choplin22a}.

Apart from the $p$-nuclides $^{136}$Ce and $^{138}$Ce, Ce is considered a primarily (main) $s-$process element, with a small $r-$process fraction. Ce was indeed observed in $s-$process enhanced stars \cite{Abia02,Aoki02}, and in the CEMP$-r/s$ stars that may exhibit $i-$process composition \cite{Denissenkov19,Choplin21,Karinkuzhi21}. Cerium is however also observed in $r-$process enhanced stars \cite{Jonsell06,Siqueira-Mello14,Holmbeck18} and its presence was reported in the kilonova (AT 2017gfo) following the gravitational wave event GW170817 \cite{Domoto22}. A large body of high-quality spectrometric data on the Galactic content of cerium has been recently collected, see \textit{e.g.} \cite{CON23} and references therein.

For a complete picture of the astronomical data of Ce, one needs to consider presolar grains \cite{Zinner14} in addition to abundances in solar material. Thermodynamic calculations show that Ce is refractory under both C-rich (C/O $>$ 1) conditions \cite{Lodders95} and under O-rich (C/O $<$ 1) conditions \cite{Lodders03}. The abundance of Ce was measured in single presolar SiC grains (which condense under C-rich conditions) of different subtypes, the source stars of which are believed to be mainly low-mass AGB stars of different metallicities and supernovae \cite{Amari95b}. Recently, the Ce abundance was measured for the first time also in a presolar silicate grain (which condense under O-rich conditions), believed to originate in a low mass AGB star \cite{Leitner22}. Also recently, Lugaro \textit{et al.} \cite{Lugaro20} proposed correlating Ce/Y ratios in Ba stars with $^{88}$Sr/$^{86}$Sr ratios in large single presolar SiC grains to infer the type of their source stars.

In this paper we report new measurements of the radiative neutron capture $(n,\gamma)$ Maxwellian-averaged cross sections (MACS) of the stable isotopes at energies relevant to the $s$ and $i$ processes; see \cite{KAP96,HAR00,AMA21} for previous measurements and  a compilation in \cite{kadonis} and references therein. 
%The motivation for remeasuring the cross sections came from the spread in past experimental results (\cite{KAP96,HAR00,AMA21}, see also \cite{kadonis} and references therein). 
Of special interest is the $(n,\gamma)$ reaction on $^{140}$Ce which is a neutron-magic ($N=82$) nucleus with a typical high neutron binding energy and small 
neutron capture cross section. $^{140}$Ce$(n,\gamma)$ feeds the potential $s-$ and 
$i-$processes branching point $^{141}$Ce (terrestrial $t_{1/2} = 32.504$ d, calculated not to change at $s-$ and $i-$process temperatures \cite{COS81,TAK87}). 
Koloczek \textit{et al. } \cite{Koloczek16} calculated that this cross section affects the production of 33 other isotopes in the $s$ process. A discrepancy between the Ce abundance observed in the
globular clusters M4 \cite{YOU03} and M22 \cite{ROE11} and that calculated by stellar models was recently pointed out by Straniero \textit{et al.} \cite{STR14} and was attributed possibly to insufficient experimental knowledge of the stellar $^{140}$Ce$(n,\gamma)$ cross section. 
%The experiment presented here was performed at 
%the Soreq Applied Research Accelerator Facility (SARAF), using the liquid-lithium target (LiLiT), an intense quasi-Maxwellian neutron source \cite{Paul19}. 

The article is organized as follows: Section II recounts the irradiation of the $^{nat}$Ce sample, Sec. III describes the detection of the radiation products by $\gamma-$spectrometry, Sec. IV details the extraction of cross sections, Sec. V is a short discussion and Sec. VI is a summary. Preliminary results of this experiment were presented in Refs \cite{SAH23,TES16}.\\

\section{\label{sec:level1} C\lowercase{e} sample irradiation}

\begin{figure}[h]
\centering
\includegraphics[width=0.85\columnwidth]{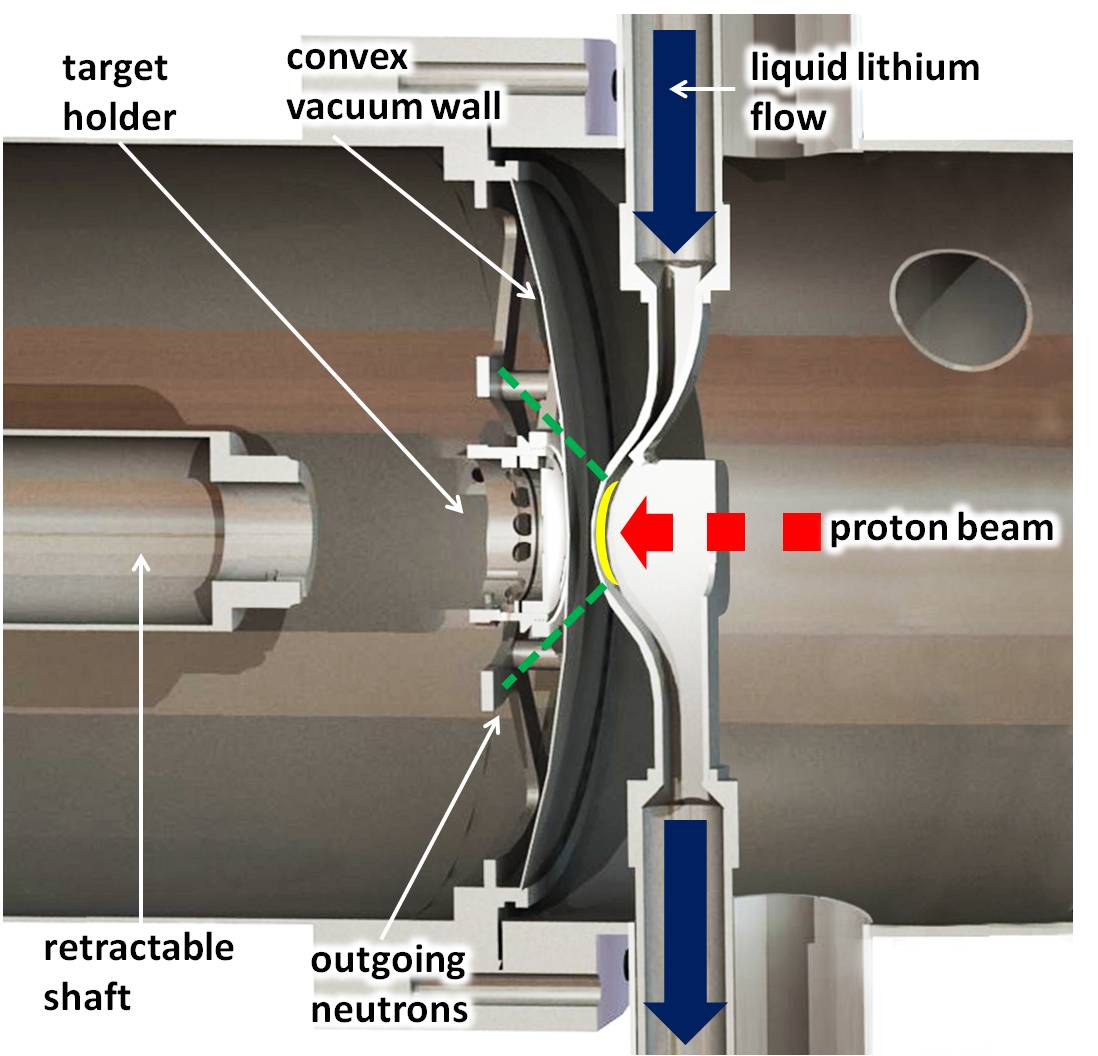}
  \caption{\label{fig:ex_setup} Cross section diagram of the Liquid-Lithium Target (LiLiT) and activation target assembly.
  The  proton beam (dashed thick arrow) impinges on the free-surface lithium film; inlet and outlet of the liquid lithium,  circulating in a closed loop at $\approx 200 ^\circ$C, are indicated by thick downward arrows 
  (see \cite{PAU19a} for details).
  The activation target sandwich (Au-Ce-Au) is mounted on a circular target holder and positioned in the outgoing neutron field (dotted lines)
  at a distance of 6-8 mm from the lithium surface in a vacuum chamber
  separated from the LiLiT chamber by a 0.5 mm stainless steel wall convex to the beam.
  The retractable shaft (at left) is used to load and unload rapidly the target assembly.}
\end{figure}

\begin{figure}[h!]
\centering
\includegraphics[width=8.7cm, trim={0.0cm 0.0cm 0.0cm 1cm}]{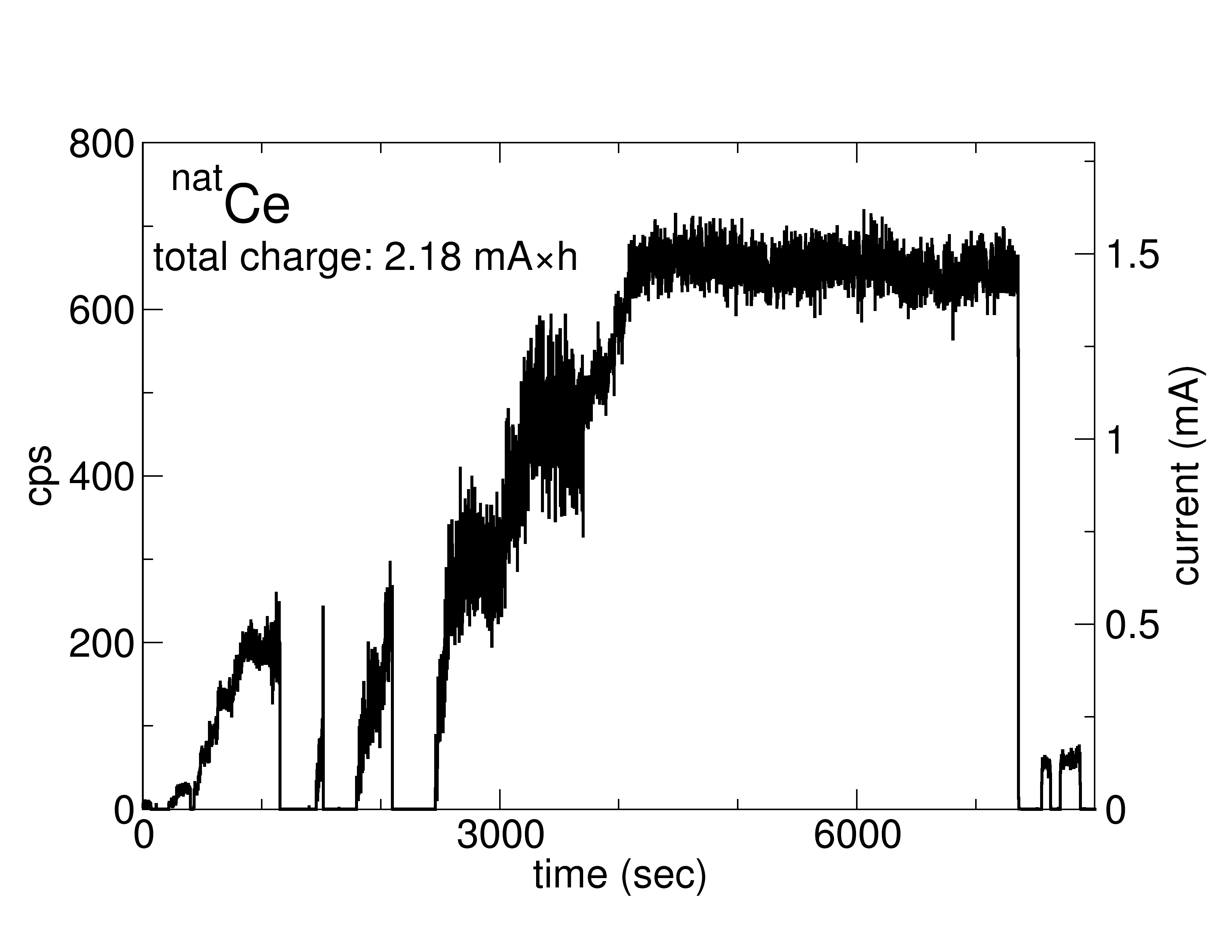}
\caption{Time profile of proton beam intensity during irradiation ($E_p = 1.924$ MeV). The left vertical axis represents the count rate of $^{235}$U fission events, produced in the fission detector (see text). The right vertical axis displays the corresponding proton beam intensity, calibrated at low intensity against an electron-suppressed Faraday cup located in front of the LiLiT chamber. The low-intensity groups near time$=0$ and time$=6750 s$ correspond to the calibration runs of the fission detector. Other gaps are periods of SARAF accelerator instability.}
\label{fig:time_profile}
\end{figure}

The neutron irradiation was performed using the Soreq Applied Research Accelerator Facility (SARAF) and the Liquid Lithium Target (LiLiT); see \cite{MAR18,PAU19a} for review articles. Quasi-maxwellian neutrons were produced by an intense proton beam (1-1.5 mA) from the SARAF superconducting linear accelerator bombarding the LiLiT target at an energy of 1.925 MeV, just above the threshold (1881 keV) of the $^{7}$Li($p,n)^7$Be reaction, following the method introduced in Ratynski \& K\"appeler \cite{RAT88}. The proton beam energy was measured by a Rutherford back scattering detector located after the accelerator modules, with a typical energy spread of 15 keV. The windowless LiLiT setup (Fig. \ref{fig:ex_setup}), which consists of liquid lithium (at $\approx 200 ^\circ$C) circulating in closed loop and producing a target film of $\approx 1.5$ mm
thickness and 18 mm wide, allows for dissipation of the beam power (2-3 kW) by fast transport 
(2-3 m/s) to a heat exchanger. Neutrons, emitted mostly in the forward direction due to the reaction kinematics, bombarded secondary activation targets located in an experimental chamber at rough vacuum separated by a 0.5 mm thick convex stainless steel wall from the LiLiT chamber and accelerator vacuum. A high-purity metallic $^{nat}$Ce target (Ce-I, 25 mm diameter, 2.114(1) g) was sandwiched  between two Au foils labeled Au(1) and Au(2) (0.110(1) g and 0.119(1) g, respectively), used as neutron fluence monitors (Table \ref{tab: isotopes}).
\begin{table}[ht]
%\begin{threeparttable}
\caption{\label{tab: isotopes} Areal density  (atoms/cm$^{2}$) of stable isotopes of cerium in targets Ce-I and Ce-II (see text), and Au(1), Au(2) targets for the above threshold irradiation.}
\begin{ruledtabular}
\begin{tabular}{lccc}
Stable Isotopes         &  Abundance & Ce-I& Ce-II       \\
$^{nat}$Ce, Au            &  ($\%$)  & (atoms/cm$^2$) &  (atoms/cm$^2$) \\
\hline 
$^{136}$Ce & 0.185(2) & 3.52(4)$\times$10$^{18}$ &  3.46(4)$\times$10$^{18}$ \\ $^{138}$Ce & 0.251(2) & 4.63(4)$\times$10$^{18}$ &  4.55(4)$\times$10$^{18}$ \\
$^{140}$Ce & 88.45(5) & 1.640(1)$\times$10$^{21}$& 1.612(1)$\times$10$^{21}$ \\
$^{142}$Ce & 11.11(5) & 2.05(1)$\times$10$^{20}$ & 2.02(1)$\times$10$^{20}$  \\
$^{197}$Au(1)   & 100    & 6.83$\times$10$^{19}$ &  7.24$\times$10$^{19}$  \\
$^{197}$Au(2)   & 100    & 7.40$\times$10$^{19}$ &  7.02$\times$10$^{19}$  \\
\end{tabular}
\end{ruledtabular}
% \begin{tablenotes}
%      \small
%      \item Uncertainties in the last significant figures are given in parentheses. The 0.4 $\%$ uncertainty has been considered in the atoms/cm$^2$ for each isotope.  
%    \end{tablenotes}
%  \end{threeparttable}
\end{table} 
Figure \ref{fig:time_profile} illustrates the time profile of the proton current during the irradiation process monitored by a commercial fission detector located in the neutron field at 0$^\circ$ relative to the incident proton beam at $\approx 80$ cm downstream (in air) of the experimental chamber. The time profile was used to calculate a correction for the decay of reaction products during the irradiation, significant for  the shortest-lived reaction product $^{137g}$Ce (t$_{1/2}=9$ h). 
\begin{figure}[h!]
\centering
\includegraphics[width=8.7cm]{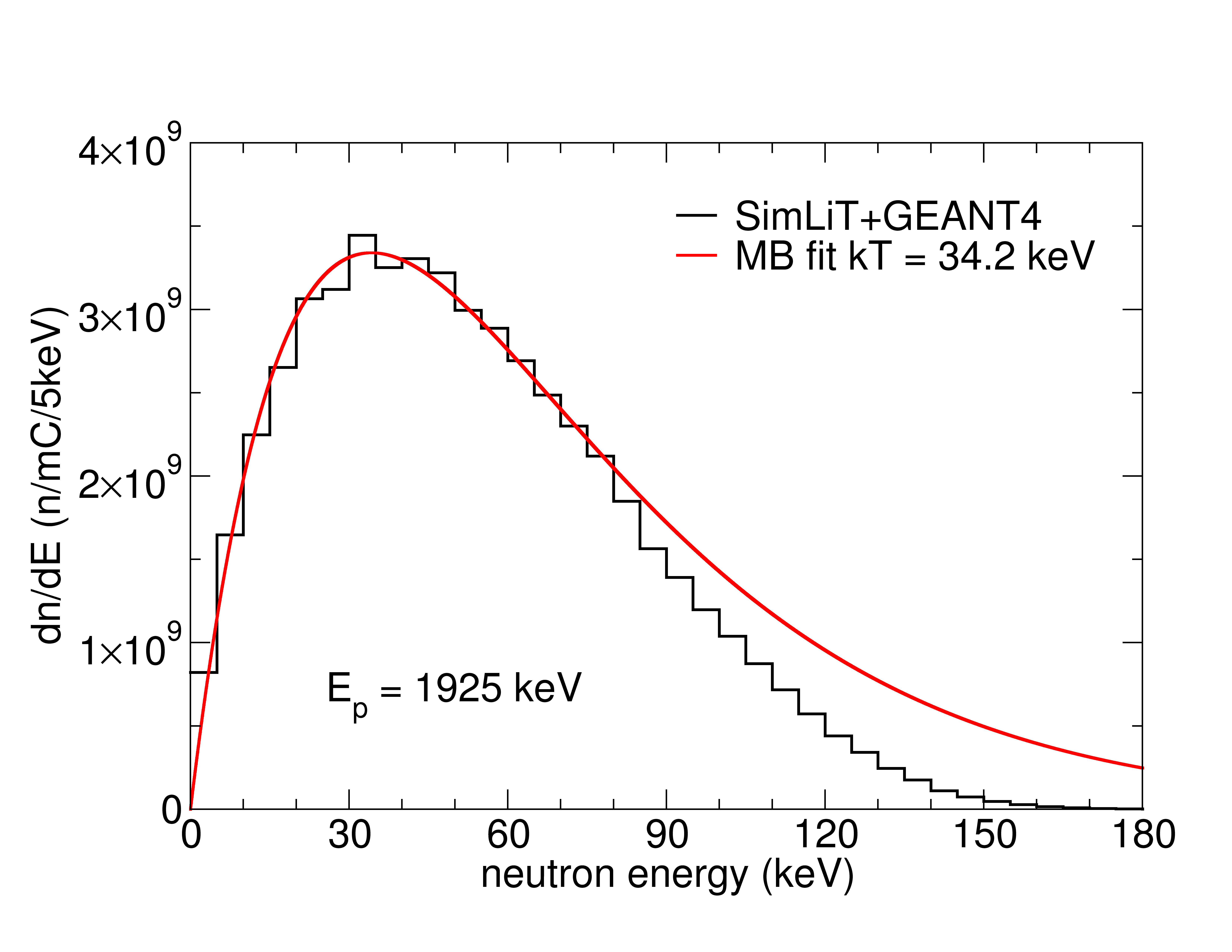}
\caption{Simulated neutron spectrum  $\frac{dn_{sim}}{dE_{n}}$  impinging on the $^{nat}$Ce- target (histogram) fitted to a Maxwell-Boltzmann flux distribution at 34.2 keV (solid line). }
\label{fig:SimLitGeant4}
\end{figure}
%, trim={0.2cm 1cm 0.7cm 0.5cm}
The energy spectrum of the neutron field seen by the target sandwich could not be experimentally determined and was calculated with the aid of the Monte Carlo code SimLiT \cite{FRI13} and the transport code GEANT4 \cite{AGO03} using a detailed model of the LiLiT chamber and target geometry. The SimLiT-GEANT4 codes were carefully benchmarked in previous experiments using the same setup \cite{TES15,PAU19a}. The simulated spectrum of the neutrons subtending the Ce target is shown in Fig. \ref{fig:SimLitGeant4} together with a Maxwell-Boltzmann flux distribution fitted to the simulated spectrum for $T=34.2$ keV. The simulated experimental spectrum (histogram in Fig. \ref{fig:SimLitGeant4}) includes 85\% of the fitted Maxwell-Boltzmann distribution at $kT$= 34.2 keV The validity of the simulated spectrum is further demonstrated by the fact that the experimental and simulated number of activated Au nuclei (denoted respectively by $N_{act}(Au)$  and $N_{act}^{ENDF}(Au)$ (see below  Eq. \eqref{eq:C_lib}) agree with each other within 1\% after normalization to proton charge. 

It had been noted in previous experiments \cite{TES15,PAU19a} that in the LiLiT setup using a thick Li target, high-energy $\gamma$ rays (17.6 and 14.6 MeV) due to the $^7$Li$(p,\gamma)^8$Be proton capture reaction, produce $(\gamma,n)$ reactions in the irradiated target which could interfere with the $(n,\gamma)$ reaction products. In the present case for example, the $^{140}$Ce$(\gamma,n)^{139}$Ce reaction can interfere with the $^{139}$Ce production via the $^{138}$Ce$(n,\gamma)^{139}$Ce investigated in this work. In order to correct quantitatively for this effect, a separate Ce target (Ce-II, Table \ref{tab: isotopes}) was irradiated with a pure $\gamma$ ray field from the $^7$Li$(p,\gamma)^8$Be reaction at a proton energy (1.810 MeV) below the neutron threshold (no neutrons present). A Au foil monitored the $\gamma$ fluence via the $^{197}$Au$(\gamma,n)^{196}$Au and $^{196}$Au(6.2 d) activity, of well-established cross section. Results of the $(n,\gamma)$ and $(\gamma,n)$ irradiations are presented in the next section.

\section{\label{sec:level2}Detection of activated $^{A+1}$C\lowercase{e} nuclei}
\begin{figure}[h!]
\centering
\includegraphics[width=8.7cm, trim={0.7cm 0cm 0.7cm 1cm}]{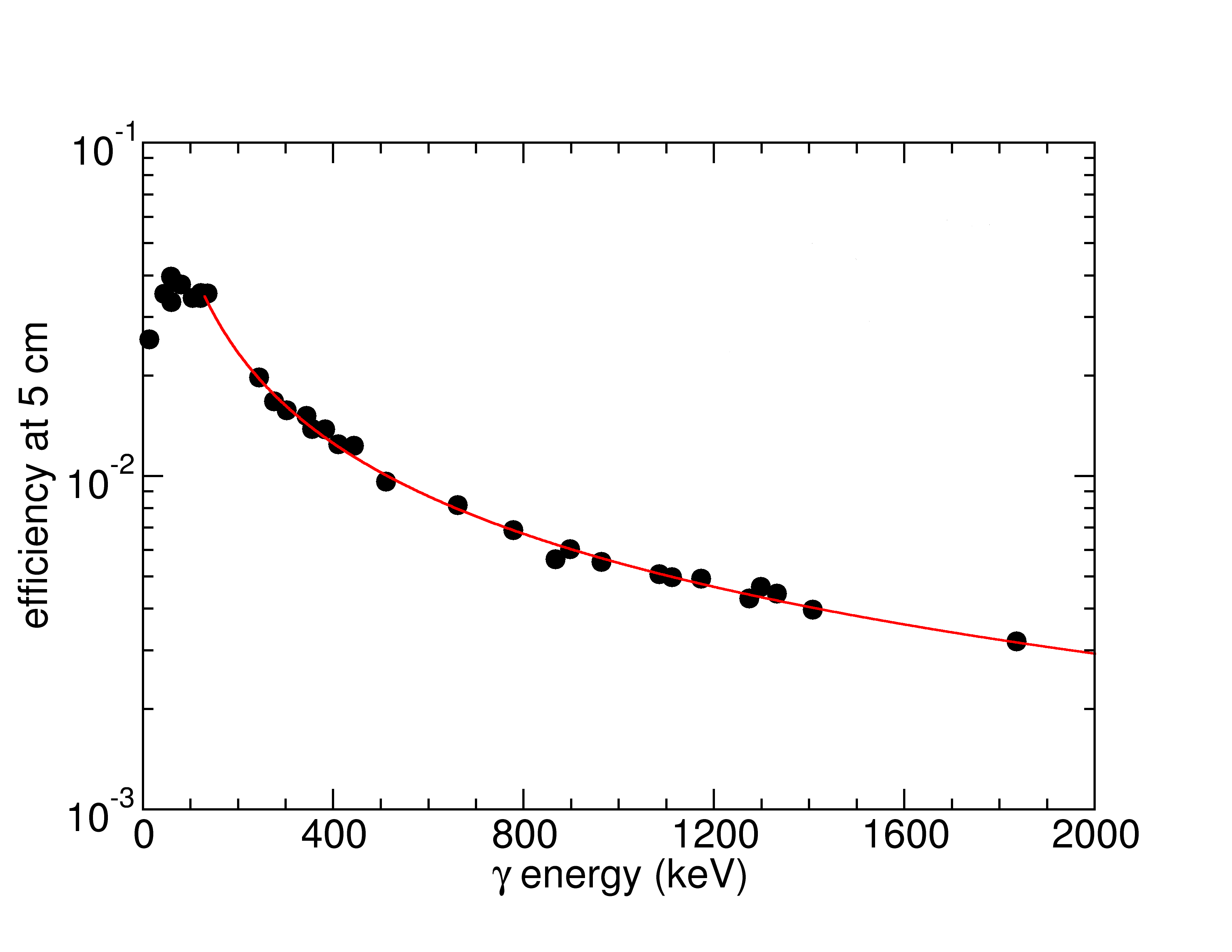}
\caption{Full-energy peak efficiency of HPGe detector using standard multi-gamma sources at 5cm apart from the end-cap. }
\label{fig: eff}
\end{figure}

\begin{figure}[h!]
\centering
\includegraphics[width=8.7cm, trim={0.0cm 1cm 0.0cm 1cm}]{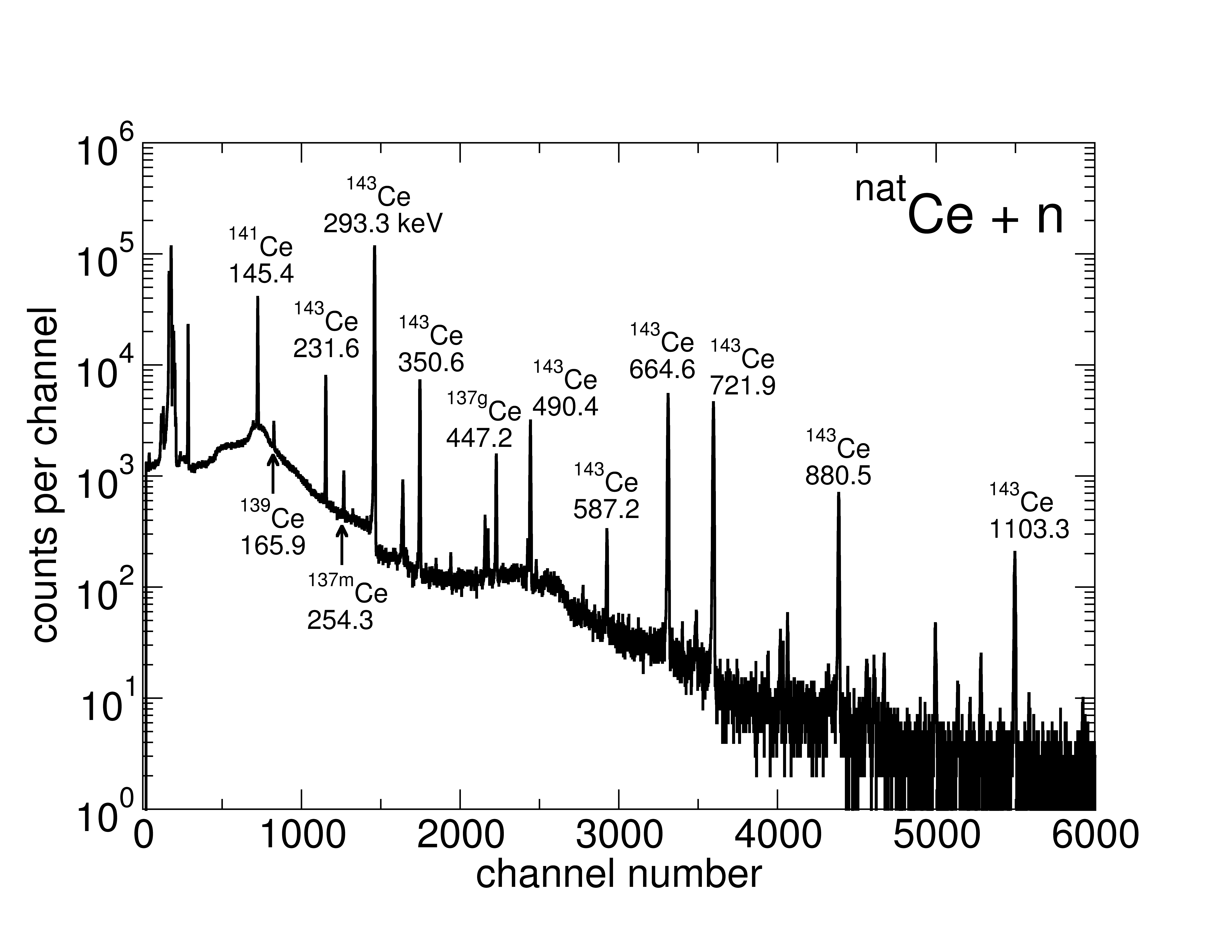}
\caption{$\gamma$-ray spectrum obtained after irradiating $^{nat}$Ce with Maxwellian neutron flux at SARAF-I above threshold energy. $\gamma$-transition of all the isotopes of $^{nat}$Ce(n,$\gamma$) reactions are identified along with the isomeric $^{137m}$Ce state. The figure is reproduced from \cite{TES16}.}
\label{fig:above_thres_irr}
\end{figure}

\begin{figure}[h!]
\centering
\vspace{20pt}
\includegraphics[width=8.7cm, trim={0.0cm 0.0cm 0.0cm 0.5cm}]{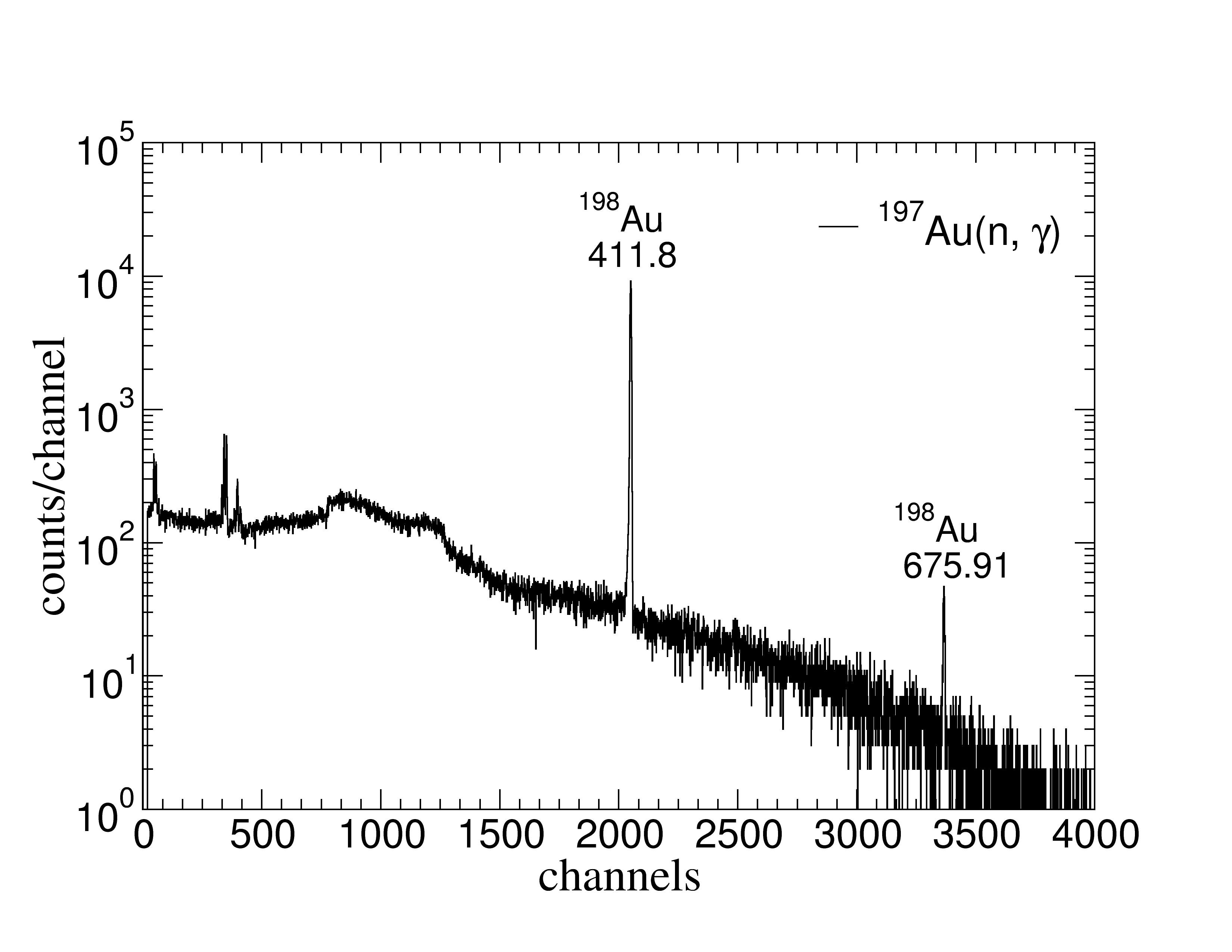}
\caption{$\gamma$-ray spectrum obtained for the upstream Au(1) monitor (above-threshold irradiation).  The $^{198}$Au transitions (411.8 keV, and 675.9 keV) are labeled.}
\label{fig:31Au_above_thres}
\end{figure}

\begin{figure}[h!]
\centering
\includegraphics[width=6cm, trim={0.0cm 0.0cm 0.0cm 0.0cm}]{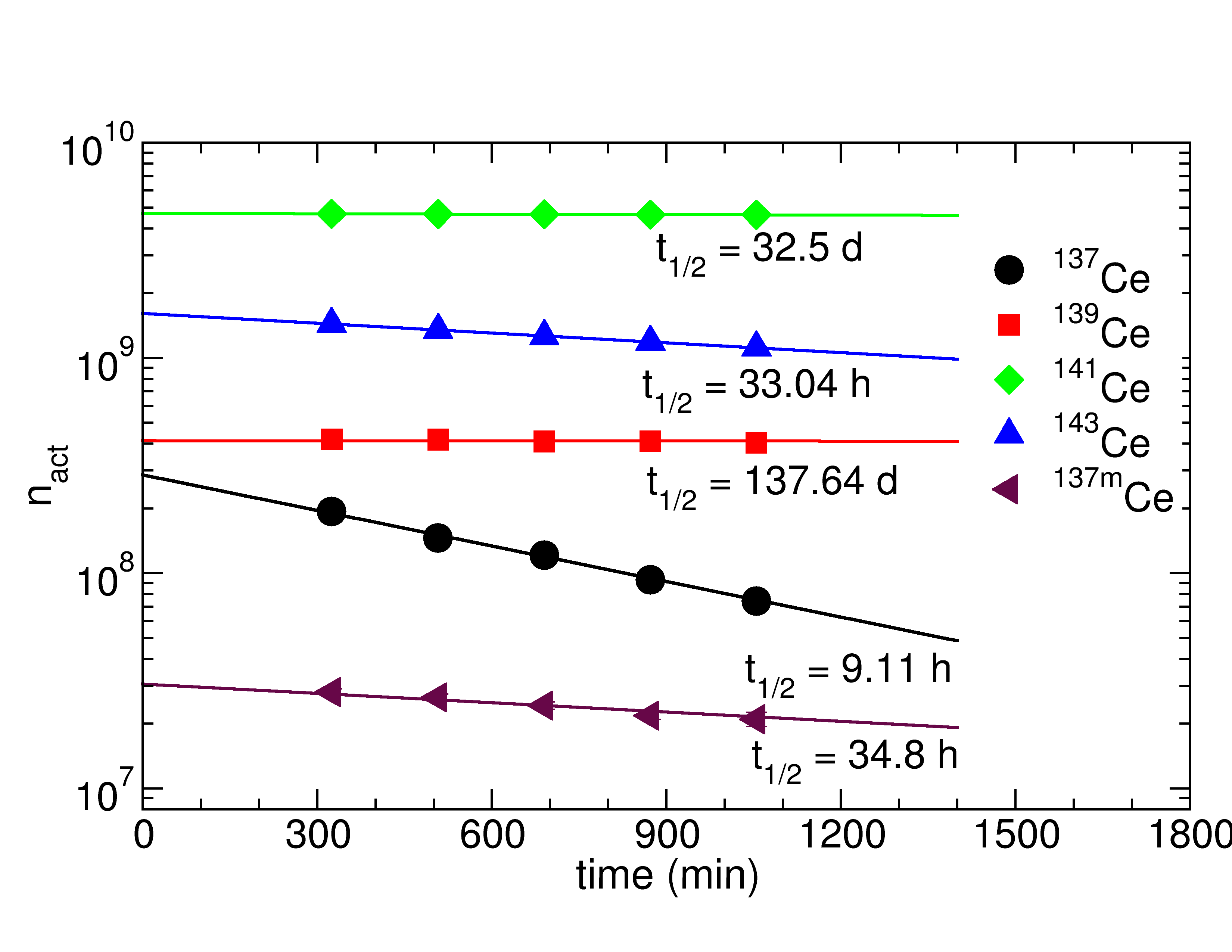}
\caption{Decay curves of $^{137,139,141,143}$Ce and $^{137m}$Ce (irradiation above neutron threshold). Data are fitted using the half-life adopted in the literature.}
\label{fig:abovethres_fit}
\end{figure}

\begin{figure}[h!]
\centering
\includegraphics[width=6cm, trim={0.0cm 0.0cm 0.0cm 1.5cm}]{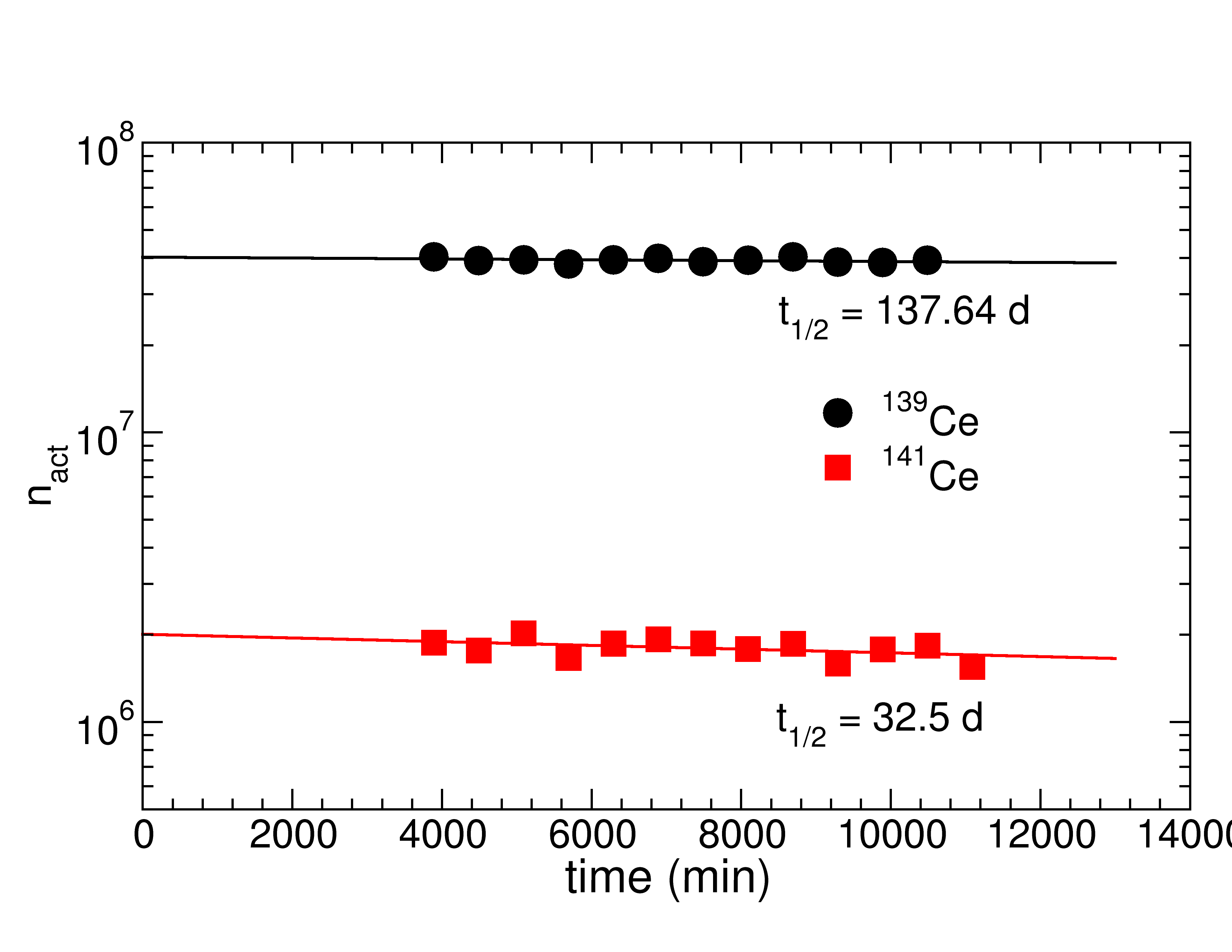}
\caption{Decay curves of $^{139,141}$Ce (irradiation under neutron threshold). Data are fitted using the half-life adopted in the literature.}
\label{fig:underthresh_fit}
\end{figure}

\begin{figure}[h]
\centering
\includegraphics[width=8.2cm, trim={0.0cm 0.0cm 0.0cm 0cm}]{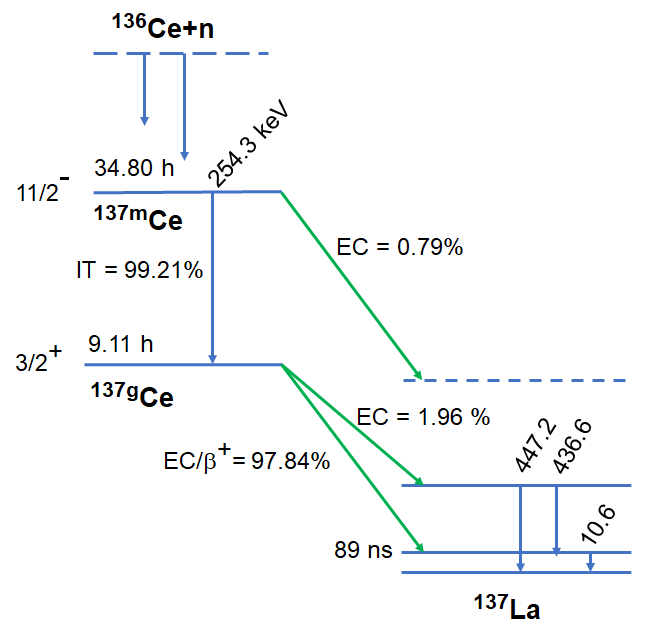}
\caption{Simplified decay scheme of $^{137}$Ce, showing the combined decay of $^{137(m+g)}$Ce. } 
\label{fig:isomeric_decay}
\end{figure}

Detailed information on the identified Ce isotopes, including their half-lives, gamma-ray transitions, and intensities are given in Table \ref{tab: isotopes_details1}. 
A shielded HPGe detector was employed to identify and measure the induced activities of reaction products. Prior to the activity measurement, the detector was calibrated, and its efficiency was determined using a standard multi-gamma radioactive source, including $^{22}$Na, $^{60}$Co, $^{88}$Y, $^{133}$Ba, $^{137}$Cs, $^{241}$Am, $^{152}$Eu, and $^{155}$Eu isotopes, positioned at a distance of 5 cm from the detector end-cap. The efficiency curve obtained from this measurement is depicted in Fig. \ref{fig: eff}. 
The spectrum of irradiated $^{nat}$Ce above neutron threshold, measured at 5 cm distance, is presented in Fig. \ref{fig:above_thres_irr}, revealing the identification of $\gamma$ lines corresponding to $^{137,139,141,143}$Ce isotopes and of a metastable state of $^{137}$Ce isotope ($^{137m}$Ce, $t_{1/2}= 34.80 h$, see Fig. \ref{fig:isomeric_decay}). Figure  \ref{fig:31Au_above_thres} represents the $\gamma-$ray spectrum Au(1) (upstream monitor) obtained in above threshold irradiation.\\  
For either irradiation (above- or under-threshold), the residual number of activated nuclei  at time $t_{cool}$ after end of irradiation is obtained using the following equation:
\begin{equation}
\label{eq:1}
n_{act}(t_{cool}) =\frac{C_{\gamma}(t_{cool})}{\epsilon_{\gamma}I_{\gamma}K_{\gamma}(1-e^{-\lambda t_{real}})} \frac{t_{real}}{t_{live}}\frac{1}{f_b};
\end{equation}
see decay curves of $n_{act}$ in Figs. \ref{fig:abovethres_fit}, \ref{fig:underthresh_fit} for above-threshold and under-threshold irradiation, respectively.
\begin{figure}[h]
\centering
\includegraphics[width=8.7cm,  trim={0.7cm 1cm 0.7cm 1cm}]{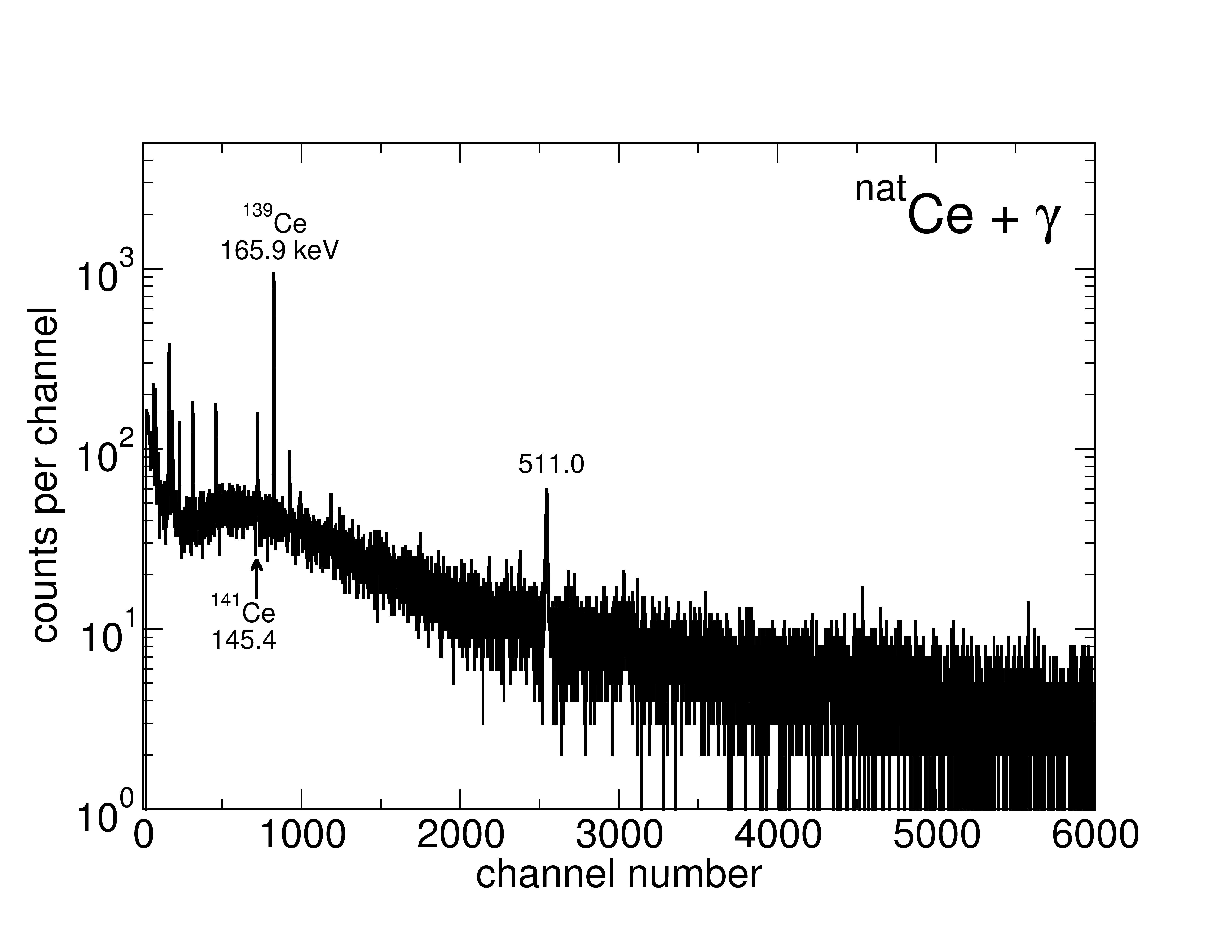}
\caption{$\gamma$ -ray spectrum obtained after irradiating  $^{nat}$Ce with $\gamma$-rays from $^7$Li(p,$\gamma$) reaction (under threshold energy). Only  $^{139}$Ce and $^{141}$Ce $\gamma$ -transitions are observed and used for correction of the $(n,\gamma)$ yields. The figure is reproduced from \cite{TES16}.}
\label{fig:under_thres}
\end{figure}
%%
%%\section{\label{sec:level1}Activated $^{A+1}$C\lowercase{e} nuclei determination}
%In both cases, in above and under threshold conditions, 

In Eq. \eqref{eq:1}, C$_\gamma (t_{cool})$ is the number of counts in the full-energy peak of a $\gamma$ line; $\epsilon_{\gamma}$, I$_\gamma$, and K$_\gamma$ are the full-energy peak efficiency (Fig. \ref{fig: eff}), $\gamma$ transition intensity and self-absorption coefficient %\cite{KGAU, KGCE} 
calculated from the target thickness data and the different $\gamma$-ray energies, respectively 
(Table \ref{tab: isotopes_details1}); $\lambda$, $t_{real}$ and $t_{live}$  are the decay constant of the transition, real and live counting time, respectively. The notation $f_b$ refers to a correction for nuclei decaying during the irradiation time, including a minor contribution for $^{137g}$Ce due to  feeding from the isomeric state $^{137m}$Ce, and is explained below.
%%%%%
%%%%%
We note here that the 447.2 keV $\gamma$-transition (Fig. \ref{fig:isomeric_decay}), which characterizes the decay of the $^{137}$Ce ground state, is practically degenerate with the 447.45 keV  (0.06\%) transition in 
$^{143}$Ce decay within the resolution of the HPGe detector. In order to correct for this contribution, the counts expected from the 447.45 keV transition were estimated as follows, relative to those of the 293.26 (42.8 $\%$)  keV $\gamma$ line in the decay of the same nucleus $^{143}$Ce:
\begin{equation}
\label{eq:count_correction}
C_{447.45} = C_{293.26}\frac{I_{447.45}}{I_{293.26}}\frac{\epsilon_{447.45}}{\epsilon_{293.26}}\frac{K_{447.45}}{K_{293.26}}.
\end{equation}
$C_{\gamma}$, $I_{\gamma}$, $\epsilon_{\gamma}$ and $K_{\gamma}$ in Eq. \eqref{eq:count_correction} represent the counts, intensity, efficiency of the detector and self-absorption coefficient of the corresponding $\gamma$ transition, respectively. The contribution of the 447.45 keV transition in $^{143}$Ce decay corresponds to $\approx$15\% of the full-energy peak observed at this energy and is subtracted from the total peak counts.\\
%\begin{table}[h!]
%\caption{\label{tab:f_b} The decay constant ($\lambda$) and corresponding beam correction factor (f$_b$) of different isotopes of stable cerium isotopes and $^{198}$Au .}
%\begin{ruledtabular}
%\begin{tabular}{lcc}
%Identified          & decay constant &  Beam corr.  \\
%isotopes     &   (s$^{-1}$)    &    factor f$_b$       \\
%\hline 
%$^{137}$Ce         &     2.113(7)$\times$10$^{-5}$    & 0.935\\

%$^{137m}$Ce         &     5.553(5)$\times$10$^{-6}$    & 0.982\\
           
%$^{139}$Ce       &   5.829(1)$\times$10$^{-8}$    & 1.000 \\
          
%$^{141}$Ce       &     2.468(1)$\times$10$^{-7}$    & 0.999 \\
                    
%$^{143}$Ce       &     5.828(2)$\times$10$^{-6}$    & 0.981\\
%$^{198}$Au           &      2.976(2)$\times$10$^{-6}$   & 0.990\\

%\end{tabular}
%\end{ruledtabular}
%\end{table} 

The fraction of activated nuclei decaying during the irradiation time is calculated via the expression
\begin{equation}
\label{eq:f_b}
f_{b} =\frac{\int_{0}^{t_{eoi}}\Phi_{n}(t) e^{-\lambda (t_{eoi} -t)}dt}{\int_{0}^{t_{b}}\Phi_{n}(t)dt}, 
\end{equation}
where $\Phi_{n}(t)$ represents the time profile of the neutron intensity on target, taken as proportional to the (neutron-induced) fission chamber yield (\textit{e.g} Fig. \ref{fig:above_thres_irr} for above-threshold irradiation). Small additional corrections, overall of the order of 1\%, need to be applied in the case of $^{137g}$Ce  to take into account feeding from the isomeric state $^{137m}$Ce  (Fig. \ref{fig:isomeric_decay}) during irradiation, cooling and counting; the expressions used for these corrections are given in the Appendix. The final values of $f_b$ are given in Table \ref{tab: isotopes_details1}.
%
%**************************************$
\begin{table}[ht]
\begin{threeparttable}
\caption{\label{tab: isotopes_details1} \small Identified isotopes from $^{nat}$Ce(n,$\gamma$) reactions, and their half-life, $\gamma$-ray transitions, and intensity \cite{NNDC}. Calculated values of the self-absorption coefficient ($K_\gamma$) and correction factor $f_b$, including $^{137m}$Ce feeding (see text), are listed.}
\begin{ruledtabular} 
\begin{tabular}{llllll}
Identified          &  Half-life     &  Detected $\gamma$- & Intensity  & K${_\gamma}$ & $f_b$\\
Isotopes    &     &  trans (keV)      &  ($\%$)   & \\
\hline 
$^{137g}$Ce       & 9.11(3) h       &      447.15(8)        & 1.68(6)&  0.987  & 0.937\\
                 &                &      436.59(9)        & 0.26(1)&   0.985   & 0.937  \\

$^{137m}$Ce      & 34.80(3) h      &      254.29(5)        & 11.1(4)& 0.966    & 0.982  \\
           
$^{139}$Ce       & 137.64(2) d  &      165.86(1)        & 79.90(13)  & 0.923   & 1.000 \\
          
$^{141}$Ce       & 32.51(1) d   &      145.443(1)       & 48.30(7) & 0.912     & 0.999  \\
                    
$^{143}$Ce       & 33.039(6) h    &      231.550(2)        &2.05(5) & 0.957    & 0.981\\
                 &                &      293.226(2)        &42.8(4) & 0.977    & 0.981\\
                 &                &      350.619(3)        &3.23(4) & 0.979    & 0.981\\
                 &                &      447.45(2)        & 0.060(3) & 0.977   & 0.981\\  
                 &                &      490.368(5)        & 2.16(3) & 0.989   & 0.981\\
                 &                &      664.57(1)         & 5.69(7) & 0.991   & 0.981\\
                 &                &      721.93(1)         & 5.39(7)  & 0.992  & 0.981\\
                 &                &      880.46(1)        & 1.031(13) & 0.994  & 0.981\\
$^{198}$Au       & 2.6941(2) d     &      411.80           &95.62     &  0.999  & 0.990\\
\end{tabular}
\end{ruledtabular}
% \begin{tablenotes}
%      \small
%      \item The half-life of $^{137g}$Ce and $^{137m}$Ce state are taken from Ref.\cite{TOR12}. 
%    \end{tablenotes}
  \end{threeparttable}
\end{table}  

The number $N_{act}^{above}$ ($N_{act}^{under}$)
of activated $^A$Ce nuclei produced during the irradiations above (under) neutron threshold (Table \ref{tab: n_act_final}) was extracted from an extrapolation of the respective decay curves to end of irradiation
(time \nolinebreak $=0$, Figs.
\ref{fig:abovethres_fit}, \ref{fig:underthresh_fit}). The net yield $N_{act}$ of the $^A$Ce$(n,\gamma)$ reactions was obtained by subtracting the yield produced by $(\gamma,n)$ reactions during the irradiation above threshold from the measured yield $N_{act}^{above}$. The subtracted yield, significant only in the case of the $^{139}$Ce product, was obtained from $N_{act}^{under}$ after normalization of the respective target thicknesses and incident proton charges (Table \ref{tab: n_act_final}). We note here that the intensity of the high-energy $\gamma$ rays produced by the $^7$Li$(p,\gamma)^8$Be reaction, responsible from the $(\gamma ,n)$ yield in our experiment, is dominated by a strong low-energy resonance at E$_R= 441$ keV \cite{ZAH95,MUN18}. The difference between the thick-target average cross section for production of high-energy $\gamma$ rays in the above-threshold ($E_p = 1.912$ MeV) and under-threshold ($E_p = 1.80$ MeV) irradiations is therefore considered negligible.

%\begin{table}[h!]
%\begin{threeparttable}
%\caption{\label{tab: Normalization_factor} Energy, target thickness, integrated current, and $^7$Li(p,$\gamma$) differential cross-section at above and under threshold irradiation.}
%\begin{ruledtabular}
%\begin{tabular}{lcccc}
%Energy               &  Ce-foil Mass & Integrated & $^7$Li(p,$\gamma_{0}$+$\gamma_{1}$)      \\
% (MeV)               &  (g)          & I (mA.hr)  & cross section($\mu$b)$^a$ \\ %\hline 
%1.92  & 2.114(1)   &   2.18(1)  &    48(5)  \\
%1.80  & 2.078(1)   &   0.57(1)   &   42(5) \\  \hline
%ratio & 1.017(1)  &   3.82(7)   &   1.2(2) 

%\end{tabular}
%\end{ruledtabular}
%%%%%%\end{table} 
%%%%%%%%%%%%%%%%%%%%%%%%%%%%%%%%%%%%%%%%%%%%%%%%%%%%%%%%%%%%%%%%%

\begin{table}[ht]
%\begin{threeparttable}
\caption{\label{tab: n_act_final} \small{Number of activated nuclei ($N_{act}$) obtained in the irradiation above neutron threshold ($N_{act}^{above}$) and under threshold ($N_{act}^{under}$) after implementing all corrections. The $(\gamma,n)$ yield during the above-threshold irradiation ($(\gamma,n)^{above}$) is obtained by normalization of $N_{act}^{under}$ for target thickness and proton charge and is subtracted from $N_{act}^{above}$ to obtain the net yield of $(n,\gamma)$ reactions ($N_{act}$).   Errors listed are statistical errors only; see Table \ref{tab: MACS_error} for the overall uncertainty budget.}}
\begin{ruledtabular}
\begin{tabular}{lclll}
%            & \multicolumn{2}{l} {Above threshold} &  \multicolumn{2}{l} {Under threshold}  \\ \hline         
\small
Reaction    & \small{${(n,\gamma)+(\gamma,n)}$}                   &  \small{$(\gamma,n)^{under}$}    &  \small{$(\gamma,n)^{above}$}  &    $(n,\gamma)$       \\
           &  $N_{act}^{above}$ &     $N_{act}^{under}$   &                         &      $N_{act}$           \\
Product    &   (10$^8$)         &  (10$^8$)                &    (10$^8$)            &    (10$^8$)           \\
\hline          
$^{137g}$Ce   & 3.04(7) &  & & 3.04(7) \\
                 
$^{137m}$Ce  &  0.310(6)  &  & &  0.310(6)\\
           
$^{139}$Ce   &  4.14(8)  & 0.403(2)& 1.57(2)  & 2.57(6)  \\
          
$^{141}$Ce   &  47.0(7)  & 0.020(1) & 0.078(2) & 46.96(7) \\
                    
$^{143}$Ce   &  16.43(4)   & &  & 16.43(4)\\
$^{198}$Au(1) &  138.8(2)  &    &    & 138.8(2)\\
$^{198}$Au(2) &  135.5(3)  &    &    & 135.5(3)\\
\end{tabular}        
\end{ruledtabular}
% \begin{tablenotes}
%     \footnotesize
%      \item $^a$ The $\gamma_{0}$+$\gamma_{1}$ cross-sections of Li(p,$\gamma$) %reactions are extracted from Ref.\cite{FIS76}.
%    \end{tablenotes}
\end{table} 

\section{\label{sec:level3}$^A$C\lowercase{e}$(n,\gamma)$ experimental cross section and 
Maxwellian-averaged cross sections}
\subsection{\label{sec:level31}Experimental cross sections}
 The net number of activated $^A$Ce nuclei $N_{act}$   can be written as 
 \begin{equation}
\label{eq:Expt_N_act}
N_{act}(x) = \sigma_{exp}(x) \phi_{total}n_{t}(x),
\end{equation}
 where $\sigma_{exp}$($\textit{x}$) is our experimental $(n,\gamma)$ cross section, sometimes termed the spectrum-averaged cross section, $\phi_{total}$ is the time-integrated neutron rate and $n_t$($\textit{x}$) is the areal density of target nuclei $x$.
In our experiment the experimental 
cross section of cerium isotopes is determined relative to that of $^{197}$Au used as monitor in the same irradiation via the equation 
 \begin{equation}
\label{eq:expt_cross}
\sigma_{exp}(x) = \sigma_{exp}(Au)\frac{n_{t}(Au)N_{act}(x)}{n_{t}(x)N_{act}(Au)}.
\end{equation}
The spectrum-averaged cross section $\sigma_{exp}(Au)$ of the  $^{197}$Au$(n,\gamma)^{198}$Au reaction is calculated via the expression
 \begin{equation}
 \sigma_{exp}(Au) =\frac{\int\sigma_{ENDF}^{Au}(E_{n})\frac{dN_{exp}}{dE_{n}}dE_{n}}{\int \frac{dN_{exp}}{dE_{n}}dE_{n}}, 
 \end{equation}
where the excitation function $\sigma_{ENDF}^{Au}(E_{n})$ is taken from the ENDF/B-VIII.0 library \cite{BRO18}. The ENDF/B-VIII.0 library was found to reproduce closely experimental  data \cite{LED11shrt,MAS14} of the $^{197}$Au$(n,\gamma)^{198}$Au reaction.
The neutron spectrum, $\frac{dN_{exp}}{dE_{n}}$, is obtained as described in Section \ref{sec:level1} from the SimLiT-GEANT4 simulation code \cite{FRI13, FEI12} for each of the Au monitors. 
The average 
cross section $\sigma_{exp}$(Au) for Au(1) is 571.9 mb and for Au(2) is 563.6 mb and the value taken in Eq. \eqref{eq:expt_cross} is 568(4) mb. The experimental 
cross sections of the $^A$Ce isotopes using Eq. \eqref{eq:expt_cross} and their uncertainty are given in Table \ref{tab: expt_cross}. 

\begin{table}[h!]
\begin{threeparttable}
\caption{\label{tab: expt_cross} Experimental neutron capture cross sections ($\sigma_{exp}$)  of the  stable isotopes of Ce.}
\begin{ruledtabular}
\begin{tabular}{lc}
Reaction                          &   Cross section \\
                                  &  ($\sigma_{exp}$) mb          \\
\hline 
$^{136}$Ce(n,$\gamma$)$^{137}$Ce  &  262(10)    \\
$^{136}$Ce(n,$\gamma$)$^{137m}$Ce &  26.7(9)   \\           
$^{138}$Ce(n,$\gamma$)$^{139}$Ce  &  163(6)    \\
$^{140}$Ce(n,$\gamma$)$^{141}$Ce  &  8.4(2)  \\
$^{142}$Ce(n,$\gamma$)$^{143}$Ce  &  23.5(9)
\end{tabular}
\end{ruledtabular}
  \end{threeparttable}
\end{table} 

\subsection{\label{sec:level4}MACS calculation}

\begin{figure}
\centering
\includegraphics[width=8.5cm, trim={0.0cm 0.0cm 0.0cm 0cm}]{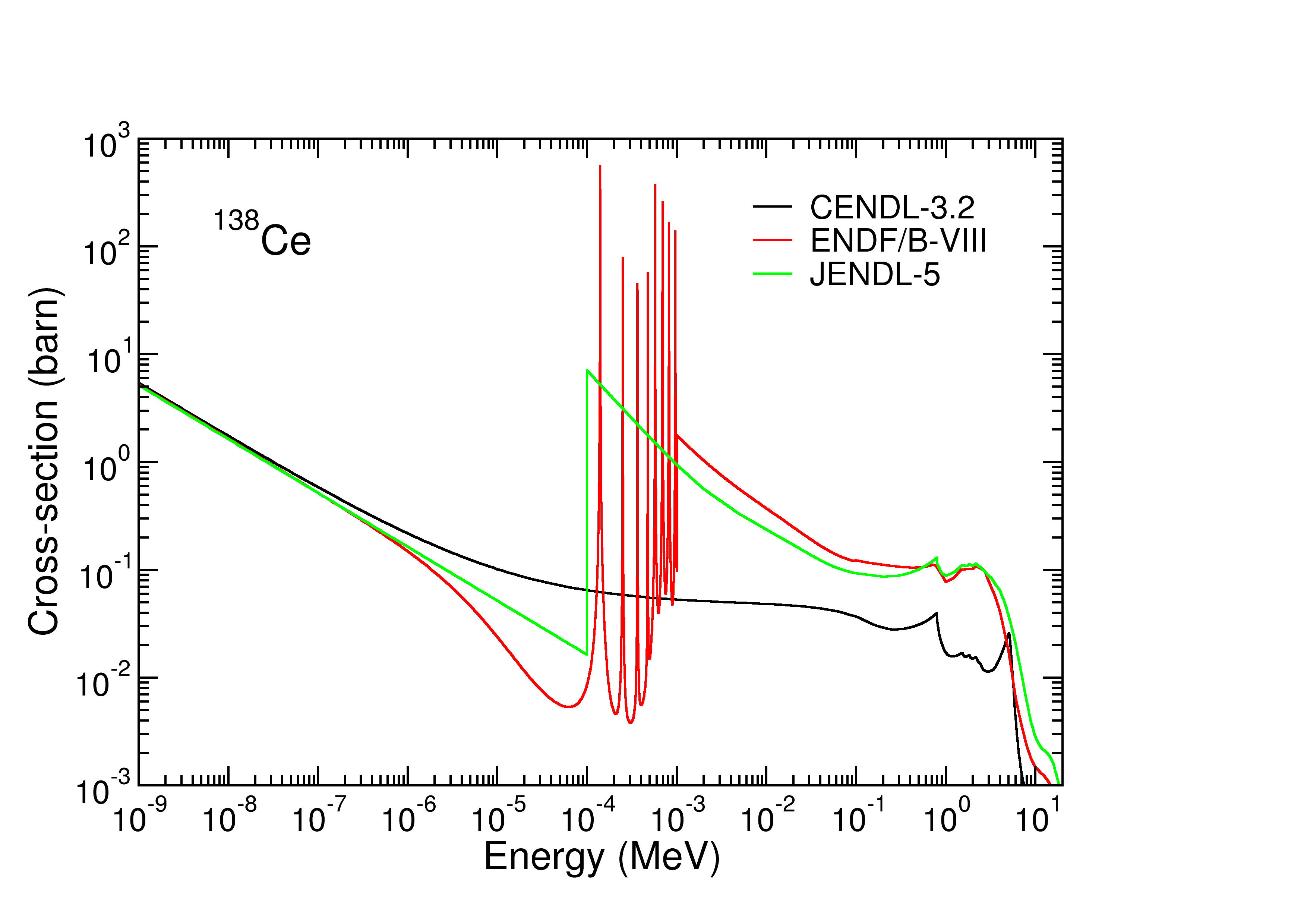}
\caption{Comparison of the $^{138}$Ce$(n,\gamma)$ excitation function evaluated in the different libraries.} 
\label{fig:lib_comparison}
\end{figure}

The 
Maxwellian$-$averaged cross section
is defined \cite{RAT88} as, 
 \begin{equation}
 MACS(kT) = 
% \frac{<\sigma v>}{v_{T}} = 
 \frac{2}{\pi} \frac{\int_{0}^{\infty}\sigma(E_{n}) E_{n} e^{(-\frac{E_{n}}{kT})}dE_{n}}{\int_{0}^{\infty} E_{n} e^{(-\frac{E_{n}}{kT})}dE_{n}}, 
\end{equation}
where $\sigma(E_{n})$ is the true energy-dependent 
$(n,\gamma$) cross section. Our measurements of the experimental cross sections  of Section \ref{sec:level31} allow us to calibrate 
%$\sigma_{lib}(E_n)$from 
evaluated neutron cross section libraries, corrected to match the measured $\sigma_{exp}$'s. Following the method described in \cite{TES15,PAU19a}, the SimLiT-GEANT4 code \cite{FRI13} is used to calculate simultaneously the numbers of $(n,\gamma)$ activated nuclei for the two gold foils, Au(1) and Au(2), and the Ce target using a 
neutron capture cross section library $\sigma_{lib}(E_n)$ and the detailed setup of our experiment. The correction factor $C_{lib}$ for $\sigma_{lib}(E_n)$ is then defined as follows:
%%%%%%%%%%%%%%%%%%%%%%%%%%%%%%%
\begin{equation}
\label{eq:C_lib}
C_{lib}= \Big[\frac{N_{act}(Ce)}{N_{act}(Au)}\Big]\Big/\Big[\frac{N_{act}^{lib}(Ce)}{N_{act}^{ENDF}(Au)}\Big],
\end{equation}
%%%%%%%%%%%%%%%%%%%%%%%%%%%%%
where $N_{act}$(Ce) and $N_{act}$ (Au) are the number of $(n,\gamma)$ activated Ce-isotopes and Au nuclei determined experimentally.  $N_{act}^{lib}$ (Ce) are the number of activated Ce-isotopes and Au nuclei determined from the SimLiT-GEANT4 simulation using the 
cross section $\sigma_{lib}(E_n)$ from a given library. The ENDF/B-VIII.0 
library was consistently used in the simulation to calculate the number of $^{198}$Au activated nuclei $N_{act}^{ENDF}$(Au).
The C$_{lib}$ values derived for several evaluated libraries are given in Table \ref{tab:C_lib_1}. Using this definition of $C_{lib}$, our experimental MACS at temperature T is then determined as 
\begin{equation}
 MACS_{lib}^{exp}(kT) = \frac{2}{\sqrt{\pi}} \frac{\int_{0}^{\infty}C_{lib}\sigma_{lib}(E_{n}) E_{n} e^{(-\frac{E_{n}}{kT})}dE_{n}}{\int_{0}^{\infty} E_{n} e^{(-\frac{E_{n}}{kT})}dE_{n}},
\end{equation} 
and the extracted values are given in Table \ref{tab: MACS_LIB} for $kT=30$ keV for the respective libraries. 
%%%%%%%%%%%%%%%%%%%%%%%%%%%%%
\begin{table}[ht]
\begin{threeparttable}
\caption{\label{tab:C_lib_1} Calculated C$_{lib}$ values (see text) using Eq. \eqref{eq:C_lib}.}
\begin{ruledtabular}
\begin{tabular}{lcccc}

Target nucleus:              &  $^{136}$Ce    & $^{138}$Ce & $^{140}$Ce & $^{142}$Ce   \\ \hline 

JENDL-5         & 1.04 & 1.23   &  1.06   &     1.29\\
JEFF-3.3        & 0.97 & 1.04   &  1.11   &     1.29\\
CENDL-3.2       & 0.85 & 3.93   &  1.06   &     1.29 \\    
ENDF/B-VIII.0   & 0.85 & 0.87   &  1.23   &     1.33\\
ROSFOND-10      & 0.85 & 0.87   &  1.23   &     1.29\\
\end{tabular}
\end{ruledtabular}
% \begin{tablenotes}
%      \footnotesize
%      \item $^a$ Note that the  C$_{lib}$ value for $^{138}$Ce using %the CENDL-3.2 data library is notably different from other values 
%(see Fig. \ref{fig:lib_comparison}.  
%    \end{tablenotes}
  \end{threeparttable}
\end{table}
Fluctuations in the $C_{lib}$ values are observed in Table \ref{tab:C_lib_1}, notably for the outlying value of $C_{CENDL}$($^{138}$Ce), reflecting different evaluated excitation functions (see Fig. \ref{fig:lib_comparison} for $^{138}$Ce showing an energy dependence of $\sigma_{CENDL}(E_n)$ different from other libraries). Nevertheless, the MACS(30 keV) values listed in Table \ref{tab: MACS_LIB} are remarkably stable; the standard deviations are taken as representing the uncertainty in the extrapolation of $\sigma_{exp}$ to a MACS at 30 keV. Table \ref{tab: MACS_error} summarizes all uncertainties involved in our experimental MACS values.\\
%%%%%%%%%%%%%%%%%%%%%%%%%%%%%
%%%%%%%%%%%%%%%%%%%%%%%%%
\begin{table}[ht]
  \begin{threeparttable}
\caption{\label{tab: MACS_LIB} Experimental MACS (mb) at 30 keV using the C$_{lib}$ values and using $(n,\gamma)$ excitation function from the different libraries and their mean values. The standard deviation, resulting from the use of different values, is taken as representing the uncertainty in the extrapolation of $\sigma_{exp}$ to a MACS at 30 keV.}
\begin{ruledtabular}
\begin{tabular}{lccccc}

Product nucleus:              &  $^{137g}$Ce &  $^{137m}$Ce   & $^{139}$Ce & $^{141}$Ce & $^{143}$Ce   \\ \hline 

JENDL-5         & 291.0     & 29.7      &181.6  &  9.60   & 25.80\\
JEFF-3.3        & 290.4     & 29.6      &181.5  &  10.5   & 25.82\\
CENDL-3.2       & 289.3     & 29.5      &185.2  &  9.61   & 25.82 \\    
ENDF/B-VIII.0   & 289.3     & 29.5      &181.6  &  9.59   & 24.90\\
ROSFOND         & 289.3     & 29.5      &181.6  &  9.59   & 25.90 \\ \hline 
Mean(std. dev.) & 290.0(8)  & 29.6(1)   &182(2) &  9.7(4) & 25.8(4) \\
\end{tabular}
\end{ruledtabular}
  \end{threeparttable}
\end{table}

\begin{table}[ht]
\begin{threeparttable}
\caption{\label{tab: MACS_error} List of uncertainties ($\%$) of MACS$_{exp}$ of isotopes $^{nat}$Ce. }
\begin{ruledtabular}
\begin{tabular}{lccccc}

 \multicolumn{5}{c}{\hspace{4cm} Uncertainty are in $\%$}    \\ \hline
Source of uncertainty    &  $^{137}$Ce       & $^{137m}$Ce & $^{138}$Ce & $^{140}$Ce & $^{142}$Ce   \\ \hline 

Target thickness         &  0.4  & 0.4   & 0.4    & 0.4    & 0.4  \\ 
Activated nuclei(Ce)     &  2.0  & 1.8   & 2.9    & 0.2    & 2.9 \\
Activated nuclei(Au)     &  0.2  & 0.2   & 0.2   & 0.2    & 0.2 \\
HPGe efficiency$^a$  &  1.5  &  1.5  &  1.5   & 1.5    & 1.5  \\
%Self absorption (K$_\gamma$)  &  0.3  &  0.3  &  0.3   & 0.6    & 0.4  \\
Intensity (I$_\gamma$)   &  0.03  &  0.03  &  $\ll$ 0.01   & $\ll$ 0.01    &  0.01  \\
Simulation$^b$                &  1.8  &  1.8  &  1.8   & 1.8    & 1.8  \\      
$\sigma_{ENDF}$(Au)      &  1.0  &  1.0  &  1.0   & 1.0    & 1.0  \\
MACS$^{exp}_{lib}$ $^c$   &  0.3  &  0.3  & 1.2    &  4.1   & 1.6 \\ \hline
Total uncertainty        & 3.3   & 3.2    & 4.1    & 4.9    &  4.2  \\      
\end{tabular}
\end{ruledtabular}
 \begin{tablenotes}
      \small
      \item $^a$ systematic error on $\gamma$ calibration sources.
      \item $^b$ includes beam energy, energy spread and geometric positioning of the activation targets.
      \item $^c$ standard deviation of MACS values from the different libraries (Table \ref{tab: MACS_LIB}). 
    \end{tablenotes}
  \end{threeparttable}
\end{table} 

Table \ref{tab: MACS}  lists our final value and overall uncertainty of the the $^ACe(n,\gamma$) 
cross section measured for all stable isotopes of cerium along with the isomeric $^{137m}$Ce state. The total $^{136}$Ce$(n,\gamma)^{137}$Ce (isomeric $+$ ground state feeding) of astrophysical significance is also listed. Table \ref{tab: MACS_kT} lists the MACS values extrapolated to a larger range of temperatures relevant to different astropysical sites, using the evaluated library JENDL-5; no uncertainties are assigned for $kT\neq 30$.
\begin{table}[ht]
\footnotesize
\begin{threeparttable}
\caption{\label{tab: MACS} Experimental MACS (mb) at 30 keV of stable isotopes of natural cerium from this work compared to values in the literature. Uncertainties include all contributions from Tables \ref{tab: n_act_final}, \ref{tab: MACS_LIB} and \ref{tab: MACS_error} }
\begin{ruledtabular}
\begin{tabular}{lcccc}
Reaction                            &  This work & \cite{KAP96} & \cite{HAR00} & \cite{kadonis}
\\  
\hline\\
$^{136}$Ce(n,$\gamma$)$^{137g}$Ce   &  290(11)  & 300(21) &  &      \\
$^{136}$Ce(n,$\gamma$)$^{137m}$Ce  &  29.6(10) & 28.2(12)&  &28.2(16)     \\  
\small$^{136}$Ce(n,$\gamma$)$^{137(g+m)}$Ce  &  320(17) &    &  & 328(21) \\ 
$^{138}$Ce(n,$\gamma$)$^{139}$Ce   &  182(8)   & 179(5)&  &  179(5)       \\
$^{140}$Ce(n,$\gamma$)$^{141}$Ce   &  9.7(5)   & 11.0(4)&  11.5(5) & 11.73(44)  \\
$^{142}$Ce(n,$\gamma$)$^{143}$Ce   &  25.8(11) & 28.3(10)&    & 29.9(10)     \\

\end{tabular}
\end{ruledtabular}
  \end{threeparttable}
\end{table} 
\begin{table}[ht]
\begin{threeparttable}
\caption{\label{tab: MACS_kT} MACS of stable Ce isotopes in mb between  $kT=10-120$ keV. JENDL-5 library is used to calculate the MACS at different $kT$ values. The MACS are listed according to the product nucleus.}
\begin{ruledtabular}
\begin{tabular}{ccccc}

 Temp(kT)&	$^{136g+m}$Ce&	$^{138}$Ce&	$^{140}$Ce&	$^{142}$Ce\\
 keV&	    MACS       & MACS       &	MACS&	MACS \\ \hline 
10	&   542  &299	   &  14.8    &	53.9  \\
20  &  	381 &215	   &  11.4    & 32.5  \\
30	&   320(17) &182(8)	   &  9.7(5)    & 25.8(11)  \\
40	&   289 &164	   &  8.47    & 22.6  \\
50  &	270 &153	   &  7.74    & 20.6  \\
60	&   258 &147	   &  7.26    & 19.3  \\
70  &	250 &142       &  6.92    & 18.4  \\
80	&   245 &139	   &  6.68    & 17.7  \\
90  &	240 &137	   &  6.51    & 17.1  \\
100 &	238 &135	   &  6.38    & 16.7  \\
110 &	236 &134	   &  6.30    & 16.3  \\
120	&   234 &134	   &  6.25    & 16.0  \\

\end{tabular}
\end{ruledtabular}
\begin{tablenotes}
      \small
      \item ---
    \end{tablenotes}
  \end{threeparttable}
\end{table}
\section{\label{sec:level6} Discussion}
Our results are in general agreement with the previous experimental studies of Kaeppeler \textit{et al.} \cite{KAP96} and Harnood \textit{et al.} \cite{HAR00} with slightly lower uncertainties on the MACS of $^{136}$Ce. The MACS value determined at 30 keV for the important case of $^{140}$Ce is lower by $\approx$15\% than reported in previous experimental studies \cite{KAP96,HAR00} and in the Kadonis data base \cite{kadonis}. This lower value partially alleviates the discrepancy highlighted by Straniero \textit{et al.} \cite{STR14}  for cerium in their study of the abundance of heavy elements in globular clusters. We note however that the resonance shape analysis done in \cite{AMA21} for an isolated $p-$wave resonance at 
$E_n=$ 5.64 keV 
leads to  a resonance strength larger than those extracted from the evaluations JENDL-5 and ENDF/B-VIII.0 pointing to a possible conflict with the trend of reduced MACS of $^{140}$Ce observed in our experiment.

\section{\label{sec:level7}Summary} 
We have determined the experimental cross section of neutron capture reactions on the stable isotopes of cerium, averaged over a quasi-Maxwellian neutron spectrum at 34.2 keV. The experiment used  the mA proton beam of the SARAF facility incident on the high-power Liquid-Lithium Target (LiLiT). In conjunction with detailed simulations of the experimental system, the experimental cross sections were extrapolated to MACS values at 30 keV and other temperatures relevant to the stellar 
$s$ process. The MACS values obtained for the important neutron-magic nucleus $^{140}$Ce is $\approx$15\% smaller than previous experimental determinations, partially resolving the discrepancy in the Ce abundance observed in  globular cluster stars.

\section{\label{sec:level8} Acknowledgments}
The SARAF and LiLiT (Soreq NRC) staff is gratefully acknowledged for their dedicated help during the experiments. We thank S. Cristallo and D. Vescovi for an enlightening discussion. This work is supported in part by the Pazy Foundation (Israel) and the German Israeli Foundation (GIF) under Grant No. I-1500-303.7/2019. MP acknowledges support by the European Union (ChETEC-INFRA). \\

\section{\label{sec:level9} Appendix}
%%%%%%%%%%%%%%%%%%%%%%%%%%%%
The $^{137m}$Ce isomer decays predominantly (99.2\%) by internal transition to the $^{137g}$Ce ground state as illustrated in Fig. \ref{fig:isomeric_decay}. The additional feeding of the ground state by decay of  the isomeric state during irradiation, cooling, and counting time leads to small corrections of the measured yield of  $^{137g}$Ce decay to obtain its  prompt feeding yield from the $(n,\gamma)$ reaction.

During irradiation,  the correction factor $f_b^{137g}$ for $^{137g}$Ce production can is expressed by

\begin{equation}
\label{eq:f_b correction}
f_{b}^{^{137g}Ce} = f_{b}+\frac{b_{f}\sigma_{m}}{\sigma_{g}}\Big[ \frac{(1-e^{-\lambda_{g}t_{b}})}{\lambda_{g}t_{b}} - \frac{(e^{-\lambda_{m}t_{b}}-e^{-\lambda_{g}t_{b}})}{(\lambda_{g}-\lambda_{m})t_{i}}\Big],
\end{equation}
where $\lambda_{g,m}$ are the respective decay constants of $^{137g,m}$Ce, $t_{b}$ and $b_{f}$ the beam irradiation time and the $m-$to$-g$ branching ratio, respectively. The correction to $f_b$ is of 0.17$\%$. 

A fraction of the $^{137m}$Ce nuclei produced during irradiation decays to the ground state during the cooling time t$_{cool}$ between end of irradiation and start of counting, given by the following expression: 
\begin{equation}
\label{eq:Isomeric_decay_cool}
f^{'}_{cool} = f_{cool}+\frac{\sigma_{m} f_{b,m}}{\sigma_{g} f_{b,g}}  \Big[\frac{e^{-\lambda_{m}t_{cool}}-e^{-\lambda_{g}t_{cool}}}{\lambda_{g}-\lambda_{m}} \Big]\lambda_{m},
\end{equation}
where $f_{cool}$ = e$^{-\lambda_m t_{cool}}$. The correction to the counted $^{137g}$Ce decays is of 0.27\%.

Similarly, a fraction of the $^{137m}$Ce states decays to the ground state during the counting time, expressed by
\begin{equation}
\vspace{0.4cm}
%\begin{align*}
\begin{split}
\label{eq:N_act correction}
N_{act}^{^{137g}Ce} = N_{act} - \frac{b_{f}N_{m}}{\lambda_{g}-\lambda_{m}} \cdot \\\Big[\frac{\lambda_{m}(e^{-\lambda_{g}t_{m}}-1)- \lambda_{g} (e^{-\lambda_{m}t_{m}}-1)}{(1-e^{-\lambda_mt_m})}\Big],
\end{split}
%\end{align*}
\end{equation}
where $N_{act}$ and $N_{m}$ are the respective number $^{137g}$Ce and  $^{137m}$Ce activated nuclei  using Eq. \eqref{eq:1} and $t_m$ represents counting time. In each measurement, the contribution of the isomeric state varies from 0.4\% to 0.7\% of the total $^{137g}$Ce population.

The corrections above  are included in the coefficients $f_b$ listed in Table \ref{tab: isotopes_details1}.\\

%\space{2 cm}

%\bibliography{EPJA_20220929}

\begin{thebibliography}{50}%
\makeatletter
\providecommand \@ifxundefined [1]{%
 \@ifx{#1\undefined}
}%
\providecommand \@ifnum [1]{%
 \ifnum #1\expandafter \@firstoftwo
 \else \expandafter \@secondoftwo
 \fi
}%
\providecommand \@ifx [1]{%
 \ifx #1\expandafter \@firstoftwo
 \else \expandafter \@secondoftwo
 \fi
}%
\providecommand \natexlab [1]{#1}%
\providecommand \enquote  [1]{``#1''}%
\providecommand \bibnamefont  [1]{#1}%
\providecommand \bibfnamefont [1]{#1}%
\providecommand \citenamefont [1]{#1}%
\providecommand \href@noop [0]{\@secondoftwo}%
\providecommand \href [0]{\begingroup \@sanitize@url \@href}%
\providecommand \@href[1]{\@@startlink{#1}\@@href}%
\providecommand \@@href[1]{\endgroup#1\@@endlink}%
\providecommand \@sanitize@url [0]{\catcode `\\12\catcode `\$12\catcode
  `\&12\catcode `\#12\catcode `\^12\catcode `\_12\catcode `\%12\relax}%
\providecommand \@@startlink[1]{}%
\providecommand \@@endlink[0]{}%
\providecommand \url  [0]{\begingroup\@sanitize@url \@url }%
\providecommand \@url [1]{\endgroup\@href {#1}{\urlprefix }}%
\providecommand \urlprefix  [0]{URL }%
\providecommand \Eprint [0]{\href }%
\providecommand \doibase [0]{http://dx.doi.org/}%
\providecommand \selectlanguage [0]{\@gobble}%
\providecommand \bibinfo  [0]{\@secondoftwo}%
\providecommand \bibfield  [0]{\@secondoftwo}%
\providecommand \translation [1]{[#1]}%
\providecommand \BibitemOpen [0]{}%
\providecommand \bibitemStop [0]{}%
\providecommand \bibitemNoStop [0]{.\EOS\space}%
\providecommand \EOS [0]{\spacefactor3000\relax}%
\providecommand \BibitemShut  [1]{\csname bibitem#1\endcsname}%
\let\auto@bib@innerbib\@empty
%</preamble>
\bibitem [{\citenamefont {Cameron}(1957)}]{CAM57}%
  \BibitemOpen
  \bibfield  {author} {\bibinfo {author} {\bibfnamefont {A.~G.~W.}\
  \bibnamefont {Cameron}},\ }\href {https://www.jstor.org/stable/40676435}
  {\bibfield  {journal} {\bibinfo  {journal} {Publ. Astron. Soc. of the
  Pacific}\ }\textbf {\bibinfo {volume} {69}},\ \bibinfo {pages} {201}
  (\bibinfo {year} {1957})}\BibitemShut {NoStop}%
\bibitem [{\citenamefont {Burbidge}\ \emph {et~al.}(1957)\citenamefont
  {Burbidge}, \citenamefont {Burbidge}, \citenamefont {Fowler},\ and\
  \citenamefont {Hoyle}}]{B2FH}%
  \BibitemOpen
  \bibfield  {author} {\bibinfo {author} {\bibfnamefont {E.~M.}\ \bibnamefont
  {Burbidge}}, \bibinfo {author} {\bibfnamefont {G.~R.}\ \bibnamefont
  {Burbidge}}, \bibinfo {author} {\bibfnamefont {W.~A.}\ \bibnamefont
  {Fowler}}, \ and\ \bibinfo {author} {\bibfnamefont {F.}~\bibnamefont
  {Hoyle}},\ }\href {\doibase https://doi.org/10.1103/RevModPhys.29.547}
  {\bibfield  {journal} {\bibinfo  {journal} {Rev. Mod. Phys.}\ }\textbf
  {\bibinfo {volume} {29}},\ \bibinfo {pages} {547} (\bibinfo {year}
  {1957})}\BibitemShut {NoStop}%
\bibitem [{\citenamefont {K\"appeler}\ \emph {et~al.}(2011)\citenamefont
  {K\"appeler}, \citenamefont {Gallino}, \citenamefont {Bisterzo},\ and\
  \citenamefont {Aoki}}]{Kappeler11}%
  \BibitemOpen
  \bibfield  {author} {\bibinfo {author} {\bibfnamefont {F.}~\bibnamefont
  {K\"appeler}}, \bibinfo {author} {\bibfnamefont {R.}~\bibnamefont {Gallino}},
  \bibinfo {author} {\bibfnamefont {S.}~\bibnamefont {Bisterzo}}, \ and\
  \bibinfo {author} {\bibfnamefont {W.}~\bibnamefont {Aoki}},\ }\href@noop {}
  {\bibfield  {journal} {\bibinfo  {journal} {Reviews of Modern Physics}\
  }\textbf {\bibinfo {volume} {83}},\ \bibinfo {pages} {157} (\bibinfo {year}
  {2011})}\BibitemShut {NoStop}%
\bibitem [{\citenamefont {Lugaro}\ \emph {et~al.}(2003)\citenamefont {Lugaro},
  \citenamefont {Herwig}, \citenamefont {Lattanzio}, \citenamefont {Gallino},\
  and\ \citenamefont {Straniero}}]{Lugaro03}%
  \BibitemOpen
  \bibfield  {author} {\bibinfo {author} {\bibfnamefont {M.}~\bibnamefont
  {Lugaro}}, \bibinfo {author} {\bibfnamefont {F.}~\bibnamefont {Herwig}},
  \bibinfo {author} {\bibfnamefont {J.~C.}\ \bibnamefont {Lattanzio}}, \bibinfo
  {author} {\bibfnamefont {R.}~\bibnamefont {Gallino}}, \ and\ \bibinfo
  {author} {\bibfnamefont {O.}~\bibnamefont {Straniero}},\ }\href@noop {}
  {\bibfield  {journal} {\bibinfo  {journal} {The Astrophysical Journal}\
  }\textbf {\bibinfo {volume} {586}},\ \bibinfo {pages} {1305} (\bibinfo {year}
  {2003})}\BibitemShut {NoStop}%
\bibitem [{\citenamefont {{Freiburghaus}}\ \emph {et~al.}(1999)\citenamefont
  {{Freiburghaus}}, \citenamefont {{Rembges}}, \citenamefont {{Rauscher}},
  \citenamefont {{Kolbe}}, \citenamefont {{Thielemann}}, \citenamefont
  {{Kratz}}, \citenamefont {{Pfeiffer}},\ and\ \citenamefont
  {{Cowan}}}]{Freiburghaus99}%
  \BibitemOpen
  \bibfield  {author} {\bibinfo {author} {\bibfnamefont {C.}~\bibnamefont
  {{Freiburghaus}}}, \bibinfo {author} {\bibfnamefont {J.~F.}\ \bibnamefont
  {{Rembges}}}, \bibinfo {author} {\bibfnamefont {T.}~\bibnamefont
  {{Rauscher}}}, \bibinfo {author} {\bibfnamefont {E.}~\bibnamefont {{Kolbe}}},
  \bibinfo {author} {\bibfnamefont {F.~K.}\ \bibnamefont {{Thielemann}}},
  \bibinfo {author} {\bibfnamefont {K.~L.}\ \bibnamefont {{Kratz}}}, \bibinfo
  {author} {\bibfnamefont {B.}~\bibnamefont {{Pfeiffer}}}, \ and\ \bibinfo
  {author} {\bibfnamefont {J.~J.}\ \bibnamefont {{Cowan}}},\ }\href@noop {}
  {\bibfield  {journal} {\bibinfo  {journal} {The Astrophysical Journal}\
  }\textbf {\bibinfo {volume} {516}},\ \bibinfo {pages} {381} (\bibinfo {year}
  {1999})}\BibitemShut {NoStop}%
\bibitem [{\citenamefont {{Cowan}}\ and\ \citenamefont
  {{Rose}}(1977)}]{Cowan77}%
  \BibitemOpen
  \bibfield  {author} {\bibinfo {author} {\bibfnamefont {J.~J.}\ \bibnamefont
  {{Cowan}}}\ and\ \bibinfo {author} {\bibfnamefont {W.~K.}\ \bibnamefont
  {{Rose}}},\ }\href@noop {} {\bibfield  {journal} {\bibinfo  {journal} {The
  Astrophysical Journal}\ }\textbf {\bibinfo {volume} {212}},\ \bibinfo {pages}
  {149} (\bibinfo {year} {1977})}\BibitemShut {NoStop}%
\bibitem [{\citenamefont {{Denissenkov}}\ \emph {et~al.}(2019)\citenamefont
  {{Denissenkov}}, \citenamefont {{Herwig}}, \citenamefont {{Woodward}},
  \citenamefont {{Andrassy}}, \citenamefont {{Pignatari}},\ and\ \citenamefont
  {{Jones}}}]{Denissenkov19}%
  \BibitemOpen
  \bibfield  {author} {\bibinfo {author} {\bibfnamefont {P.~A.}\ \bibnamefont
  {{Denissenkov}}}, \bibinfo {author} {\bibfnamefont {F.}~\bibnamefont
  {{Herwig}}}, \bibinfo {author} {\bibfnamefont {P.}~\bibnamefont
  {{Woodward}}}, \bibinfo {author} {\bibfnamefont {R.}~\bibnamefont
  {{Andrassy}}}, \bibinfo {author} {\bibfnamefont {M.}~\bibnamefont
  {{Pignatari}}}, \ and\ \bibinfo {author} {\bibfnamefont {S.}~\bibnamefont
  {{Jones}}},\ }\href@noop {} {\bibfield  {journal} {\bibinfo  {journal}
  {Monthly Notices of the Royal Astronomical Society}\ }\textbf {\bibinfo
  {volume} {488}},\ \bibinfo {pages} {4258} (\bibinfo {year}
  {2019})}\BibitemShut {NoStop}%
\bibitem [{\citenamefont {{Choplin}}\ \emph {et~al.}(2021)\citenamefont
  {{Choplin}}, \citenamefont {{Siess}},\ and\ \citenamefont
  {{Goriely}}}]{Choplin21}%
  \BibitemOpen
  \bibfield  {author} {\bibinfo {author} {\bibfnamefont {A.}~\bibnamefont
  {{Choplin}}}, \bibinfo {author} {\bibfnamefont {L.}~\bibnamefont {{Siess}}},
  \ and\ \bibinfo {author} {\bibfnamefont {S.}~\bibnamefont {{Goriely}}},\
  }\href@noop {} {\bibfield  {journal} {\bibinfo  {journal} {Astronomy \&
  Astrophysics}\ }\textbf {\bibinfo {volume} {648}},\ \bibinfo {pages} {A119}
  (\bibinfo {year} {2021})}\BibitemShut {NoStop}%
\bibitem [{\citenamefont {{Choplin}}\ \emph {et~al.}(2022)\citenamefont
  {{Choplin}}, \citenamefont {{Siess}},\ and\ \citenamefont
  {{Goriely}}}]{Choplin22b}%
  \BibitemOpen
  \bibfield  {author} {\bibinfo {author} {\bibfnamefont {A.}~\bibnamefont
  {{Choplin}}}, \bibinfo {author} {\bibfnamefont {L.}~\bibnamefont {{Siess}}},
  \ and\ \bibinfo {author} {\bibfnamefont {S.}~\bibnamefont {{Goriely}}},\
  }\href@noop {} {\bibfield  {journal} {\bibinfo  {journal} {Astronomy \&
  Astrophysics}\ }\textbf {\bibinfo {volume} {667}},\ \bibinfo {pages} {A155}
  (\bibinfo {year} {2022})}\BibitemShut {NoStop}%
\bibitem [{\citenamefont {Meyer}(1994)}]{Meyer94}%
  \BibitemOpen
  \bibfield  {author} {\bibinfo {author} {\bibfnamefont {B.~S.}\ \bibnamefont
  {Meyer}},\ }\href@noop {} {\bibfield  {journal} {\bibinfo  {journal} {Annual
  Review of Astronomy and Astrophysics}\ }\textbf {\bibinfo {volume} {32}},\
  \bibinfo {pages} {153} (\bibinfo {year} {1994})}\BibitemShut {NoStop}%
\bibitem [{\citenamefont {{Choplin, A.}}\ \emph {et~al.}(2022)\citenamefont
  {{Choplin, A.}}, \citenamefont {{Goriely, S.}}, \citenamefont {{Hirschi,
  R.}}, \citenamefont {{Tominaga, N.}},\ and\ \citenamefont {{Meynet,
  G.}}}]{Choplin22a}%
  \BibitemOpen
  \bibfield  {author} {\bibinfo {author} {\bibnamefont {{Choplin, A.}}},
  \bibinfo {author} {\bibnamefont {{Goriely, S.}}}, \bibinfo {author}
  {\bibnamefont {{Hirschi, R.}}}, \bibinfo {author} {\bibnamefont {{Tominaga,
  N.}}}, \ and\ \bibinfo {author} {\bibnamefont {{Meynet, G.}}},\ }\href
  {\doibase 10.1051/0004-6361/202243331} {\bibfield  {journal} {\bibinfo
  {journal} {A\&A}\ }\textbf {\bibinfo {volume} {661}},\ \bibinfo {pages} {A86}
  (\bibinfo {year} {2022})}\BibitemShut {NoStop}%
\bibitem [{\citenamefont {Abia}\ \emph {et~al.}(2002)\citenamefont {Abia},
  \citenamefont {Domínguez}, \citenamefont {Gallino}, \citenamefont {Busso},
  \citenamefont {Masera}, \citenamefont {Straniero}, \citenamefont
  {de~Laverny}, \citenamefont {Plez},\ and\ \citenamefont {Isern}}]{Abia02}%
  \BibitemOpen
  \bibfield  {author} {\bibinfo {author} {\bibfnamefont {C.}~\bibnamefont
  {Abia}}, \bibinfo {author} {\bibfnamefont {I.}~\bibnamefont {Domínguez}},
  \bibinfo {author} {\bibfnamefont {R.}~\bibnamefont {Gallino}}, \bibinfo
  {author} {\bibfnamefont {M.}~\bibnamefont {Busso}}, \bibinfo {author}
  {\bibfnamefont {S.}~\bibnamefont {Masera}}, \bibinfo {author} {\bibfnamefont
  {O.}~\bibnamefont {Straniero}}, \bibinfo {author} {\bibfnamefont
  {P.}~\bibnamefont {de~Laverny}}, \bibinfo {author} {\bibfnamefont
  {B.}~\bibnamefont {Plez}}, \ and\ \bibinfo {author} {\bibfnamefont
  {J.}~\bibnamefont {Isern}},\ }\href@noop {} {\bibfield  {journal} {\bibinfo
  {journal} {The Astrophysical Journal}\ }\textbf {\bibinfo {volume} {579}},\
  \bibinfo {pages} {817} (\bibinfo {year} {2002})}\BibitemShut {NoStop}%
\bibitem [{\citenamefont {Aoki}\ \emph {et~al.}(2002)\citenamefont {Aoki},
  \citenamefont {Ryan}, \citenamefont {Norris}, \citenamefont {Beers},
  \citenamefont {Ando},\ and\ \citenamefont {Tsangarides}}]{Aoki02}%
  \BibitemOpen
  \bibfield  {author} {\bibinfo {author} {\bibfnamefont {W.}~\bibnamefont
  {Aoki}}, \bibinfo {author} {\bibfnamefont {S.~G.}\ \bibnamefont {Ryan}},
  \bibinfo {author} {\bibfnamefont {J.~E.}\ \bibnamefont {Norris}}, \bibinfo
  {author} {\bibfnamefont {T.~C.}\ \bibnamefont {Beers}}, \bibinfo {author}
  {\bibfnamefont {H.}~\bibnamefont {Ando}}, \ and\ \bibinfo {author}
  {\bibfnamefont {S.}~\bibnamefont {Tsangarides}},\ }\href@noop {} {\bibfield
  {journal} {\bibinfo  {journal} {The Astrophysical Journal}\ }\textbf
  {\bibinfo {volume} {580}},\ \bibinfo {pages} {1149} (\bibinfo {year}
  {2002})}\BibitemShut {NoStop}%
\bibitem [{\citenamefont {{Karinkuzhi}}\ \emph {et~al.}(2021)\citenamefont
  {{Karinkuzhi}}, \citenamefont {{Van Eck}}, \citenamefont {{Goriely}},
  \citenamefont {{Siess}}, \citenamefont {{Jorissen}}, \citenamefont {{Merle}},
  \citenamefont {{Escorza}},\ and\ \citenamefont {{Masseron}}}]{Karinkuzhi21}%
  \BibitemOpen
  \bibfield  {author} {\bibinfo {author} {\bibfnamefont {D.}~\bibnamefont
  {{Karinkuzhi}}}, \bibinfo {author} {\bibfnamefont {S.}~\bibnamefont {{Van
  Eck}}}, \bibinfo {author} {\bibfnamefont {S.}~\bibnamefont {{Goriely}}},
  \bibinfo {author} {\bibfnamefont {L.}~\bibnamefont {{Siess}}}, \bibinfo
  {author} {\bibfnamefont {A.}~\bibnamefont {{Jorissen}}}, \bibinfo {author}
  {\bibfnamefont {T.}~\bibnamefont {{Merle}}}, \bibinfo {author} {\bibfnamefont
  {A.}~\bibnamefont {{Escorza}}}, \ and\ \bibinfo {author} {\bibfnamefont
  {T.}~\bibnamefont {{Masseron}}},\ }\href@noop {} {\bibfield  {journal}
  {\bibinfo  {journal} {Astronomy \& Astrophysics}\ }\textbf {\bibinfo {volume}
  {645}},\ \bibinfo {pages} {A61} (\bibinfo {year} {2021})}\BibitemShut
  {NoStop}%
\bibitem [{\citenamefont {{Jonsell}}\ \emph {et~al.}(2006)\citenamefont
  {{Jonsell}}, \citenamefont {{Barklem}}, \citenamefont {{Gustafsson}},
  \citenamefont {{Christlieb}}, \citenamefont {{Hill}}, \citenamefont
  {{Beers}},\ and\ \citenamefont {{Holmberg}}}]{Jonsell06}%
  \BibitemOpen
  \bibfield  {author} {\bibinfo {author} {\bibfnamefont {K.}~\bibnamefont
  {{Jonsell}}}, \bibinfo {author} {\bibfnamefont {P.~S.}\ \bibnamefont
  {{Barklem}}}, \bibinfo {author} {\bibfnamefont {B.}~\bibnamefont
  {{Gustafsson}}}, \bibinfo {author} {\bibfnamefont {N.}~\bibnamefont
  {{Christlieb}}}, \bibinfo {author} {\bibfnamefont {V.}~\bibnamefont
  {{Hill}}}, \bibinfo {author} {\bibfnamefont {T.~C.}\ \bibnamefont {{Beers}}},
  \ and\ \bibinfo {author} {\bibfnamefont {J.}~\bibnamefont {{Holmberg}}},\
  }\href@noop {} {\bibfield  {journal} {\bibinfo  {journal} {Astronomy \&
  Astrophysics}\ }\textbf {\bibinfo {volume} {451}},\ \bibinfo {pages} {651}
  (\bibinfo {year} {2006})}\BibitemShut {NoStop}%
\bibitem [{\citenamefont {{Siqueira Mello}}\ \emph {et~al.}(2014)\citenamefont
  {{Siqueira Mello}}, \citenamefont {{Hill}}, \citenamefont {{Barbuy}},
  \citenamefont {{Spite}}, \citenamefont {{Spite}}, \citenamefont {{Beers}},
  \citenamefont {{Caffau}}, \citenamefont {{Bonifacio}}, \citenamefont
  {{Cayrel}}, \citenamefont {{Fran{\c{c}}ois}}, \citenamefont {{Schatz}},\ and\
  \citenamefont {{Wanajo}}}]{Siqueira-Mello14}%
  \BibitemOpen
  \bibfield  {author} {\bibinfo {author} {\bibfnamefont {C.}~\bibnamefont
  {{Siqueira Mello}}}, \bibinfo {author} {\bibfnamefont {V.}~\bibnamefont
  {{Hill}}}, \bibinfo {author} {\bibfnamefont {B.}~\bibnamefont {{Barbuy}}},
  \bibinfo {author} {\bibfnamefont {M.}~\bibnamefont {{Spite}}}, \bibinfo
  {author} {\bibfnamefont {F.}~\bibnamefont {{Spite}}}, \bibinfo {author}
  {\bibfnamefont {T.~C.}\ \bibnamefont {{Beers}}}, \bibinfo {author}
  {\bibfnamefont {E.}~\bibnamefont {{Caffau}}}, \bibinfo {author}
  {\bibfnamefont {P.}~\bibnamefont {{Bonifacio}}}, \bibinfo {author}
  {\bibfnamefont {R.}~\bibnamefont {{Cayrel}}}, \bibinfo {author}
  {\bibfnamefont {P.}~\bibnamefont {{Fran{\c{c}}ois}}}, \bibinfo {author}
  {\bibfnamefont {H.}~\bibnamefont {{Schatz}}}, \ and\ \bibinfo {author}
  {\bibfnamefont {S.}~\bibnamefont {{Wanajo}}},\ }\href@noop {} {\bibfield
  {journal} {\bibinfo  {journal} {Astronomy \& Astrophysics}\ }\textbf
  {\bibinfo {volume} {565}},\ \bibinfo {pages} {A93} (\bibinfo {year}
  {2014})}\BibitemShut {NoStop}%
\bibitem [{\citenamefont {{Holmbeck}}\ \emph {et~al.}(2018)\citenamefont
  {{Holmbeck}}, \citenamefont {{Beers}}, \citenamefont {{Roederer}},
  \citenamefont {{Placco}}, \citenamefont {{Hansen}}, \citenamefont {{Sakari}},
  \citenamefont {{Sneden}}, \citenamefont {{Liu}}, \citenamefont {{Lee}},
  \citenamefont {{Cowan}},\ and\ \citenamefont {{Frebel}}}]{Holmbeck18}%
  \BibitemOpen
  \bibfield  {author} {\bibinfo {author} {\bibfnamefont {E.~M.}\ \bibnamefont
  {{Holmbeck}}}, \bibinfo {author} {\bibfnamefont {T.~C.}\ \bibnamefont
  {{Beers}}}, \bibinfo {author} {\bibfnamefont {I.~U.}\ \bibnamefont
  {{Roederer}}}, \bibinfo {author} {\bibfnamefont {V.~M.}\ \bibnamefont
  {{Placco}}}, \bibinfo {author} {\bibfnamefont {T.~T.}\ \bibnamefont
  {{Hansen}}}, \bibinfo {author} {\bibfnamefont {C.~M.}\ \bibnamefont
  {{Sakari}}}, \bibinfo {author} {\bibfnamefont {C.}~\bibnamefont {{Sneden}}},
  \bibinfo {author} {\bibfnamefont {C.}~\bibnamefont {{Liu}}}, \bibinfo
  {author} {\bibfnamefont {Y.~S.}\ \bibnamefont {{Lee}}}, \bibinfo {author}
  {\bibfnamefont {J.~J.}\ \bibnamefont {{Cowan}}}, \ and\ \bibinfo {author}
  {\bibfnamefont {A.}~\bibnamefont {{Frebel}}},\ }\href@noop {} {\bibfield
  {journal} {\bibinfo  {journal} {The Astrophysical Journal Letters}\ }\textbf
  {\bibinfo {volume} {859}},\ \bibinfo {pages} {L24} (\bibinfo {year}
  {2018})}\BibitemShut {NoStop}%
\bibitem [{\citenamefont {{Domoto}}\ \emph {et~al.}(2022)\citenamefont
  {{Domoto}}, \citenamefont {{Tanaka}}, \citenamefont {{Kato}}, \citenamefont
  {{Kawaguchi}}, \citenamefont {{Hotokezaka}},\ and\ \citenamefont
  {{Wanajo}}}]{Domoto22}%
  \BibitemOpen
  \bibfield  {author} {\bibinfo {author} {\bibfnamefont {N.}~\bibnamefont
  {{Domoto}}}, \bibinfo {author} {\bibfnamefont {M.}~\bibnamefont {{Tanaka}}},
  \bibinfo {author} {\bibfnamefont {D.}~\bibnamefont {{Kato}}}, \bibinfo
  {author} {\bibfnamefont {K.}~\bibnamefont {{Kawaguchi}}}, \bibinfo {author}
  {\bibfnamefont {K.}~\bibnamefont {{Hotokezaka}}}, \ and\ \bibinfo {author}
  {\bibfnamefont {S.}~\bibnamefont {{Wanajo}}},\ }\href@noop {} {\bibfield
  {journal} {\bibinfo  {journal} {The Astrophysical Journal}\ }\textbf
  {\bibinfo {volume} {939}},\ \bibinfo {pages} {8} (\bibinfo {year}
  {2022})}\BibitemShut {NoStop}%
\bibitem [{\citenamefont {{Contursi, G.}}\ \emph {et~al.}(2023)\citenamefont
  {{Contursi, G.}}, \citenamefont {{de Laverny, P.}}, \citenamefont
  {{Recio-Blanco, A.}}, \citenamefont {{Spitoni, E.}}, \citenamefont {{Palicio,
  P. A.}}, \citenamefont {{Poggio, E.}}, \citenamefont {{Grisoni, V.}},
  \citenamefont {{Cescutti, G.}}, \citenamefont {{Matteucci, F.}},
  \citenamefont {{Spina, L.}}, \citenamefont {{\'Alvarez, M. A.}},
  \citenamefont {{Kordopatis, G.}}, \citenamefont {{Ordenovic, C.}},
  \citenamefont {{Oreshina-Slezak, I.}},\ and\ \citenamefont {{Zhao,
  H.}}}]{CON23}%
  \BibitemOpen
  \bibfield  {author} {\bibinfo {author} {\bibnamefont {{Contursi, G.}}},
  \bibinfo {author} {\bibnamefont {{de Laverny, P.}}}, \bibinfo {author}
  {\bibnamefont {{Recio-Blanco, A.}}}, \bibinfo {author} {\bibnamefont
  {{Spitoni, E.}}}, \bibinfo {author} {\bibnamefont {{Palicio, P. A.}}},
  \bibinfo {author} {\bibnamefont {{Poggio, E.}}}, \bibinfo {author}
  {\bibnamefont {{Grisoni, V.}}}, \bibinfo {author} {\bibnamefont {{Cescutti,
  G.}}}, \bibinfo {author} {\bibnamefont {{Matteucci, F.}}}, \bibinfo {author}
  {\bibnamefont {{Spina, L.}}}, \bibinfo {author} {\bibnamefont {{\'Alvarez, M.
  A.}}}, \bibinfo {author} {\bibnamefont {{Kordopatis, G.}}}, \bibinfo {author}
  {\bibnamefont {{Ordenovic, C.}}}, \bibinfo {author} {\bibnamefont
  {{Oreshina-Slezak, I.}}}, \ and\ \bibinfo {author} {\bibnamefont {{Zhao,
  H.}}},\ }\href {\doibase 10.1051/0004-6361/202244469} {\bibfield  {journal}
  {\bibinfo  {journal} {A\&A}\ }\textbf {\bibinfo {volume} {670}},\ \bibinfo
  {pages} {A106} (\bibinfo {year} {2023})}\BibitemShut {NoStop}%
\bibitem [{\citenamefont {{Zinner}}(2014)}]{Zinner14}%
  \BibitemOpen
  \bibfield  {author} {\bibinfo {author} {\bibfnamefont {E.}~\bibnamefont
  {{Zinner}}},\ }in\ \href@noop {} {\emph {\bibinfo {booktitle} {Meteorites and
  Cosmochemical Processes}}},\ Vol.~\bibinfo {volume} {1},\ \bibinfo {editor}
  {edited by\ \bibinfo {editor} {\bibfnamefont {A.~M.}\ \bibnamefont
  {{Davis}}}}\ (\bibinfo {year} {2014})\ pp.\ \bibinfo {pages}
  {181-213}\BibitemShut {NoStop}%
\bibitem [{\citenamefont {{Lodders}}\ and\ \citenamefont
  {{Fegley}}(1995)}]{Lodders95}%
  \BibitemOpen
  \bibfield  {author} {\bibinfo {author} {\bibfnamefont {K.}~\bibnamefont
  {{Lodders}}}\ and\ \bibinfo {author} {\bibfnamefont {J.}~\bibnamefont
  {{Fegley}}, \bibfnamefont {B.}},\ }\href@noop {} {\bibfield  {journal}
  {\bibinfo  {journal} {Meteoritics}\ }\textbf {\bibinfo {volume} {30}},\
  \bibinfo {pages} {661} (\bibinfo {year} {1995})}\BibitemShut {NoStop}%
\bibitem [{\citenamefont {{Lodders}}(2003)}]{Lodders03}%
  \BibitemOpen
  \bibfield  {author} {\bibinfo {author} {\bibfnamefont {K.}~\bibnamefont
  {{Lodders}}},\ }\href@noop {} {\bibfield  {journal} {\bibinfo  {journal} {The
  Astrophysical Journal}\ }\textbf {\bibinfo {volume} {591}},\ \bibinfo {pages}
  {1220} (\bibinfo {year} {2003})}\BibitemShut {NoStop}%
\bibitem [{\citenamefont {{Amari}}\ \emph {et~al.}(1995)\citenamefont
  {{Amari}}, \citenamefont {{Hoppe}}, \citenamefont {{Zinner}},\ and\
  \citenamefont {{Lewis}}}]{Amari95b}%
  \BibitemOpen
  \bibfield  {author} {\bibinfo {author} {\bibfnamefont {S.}~\bibnamefont
  {{Amari}}}, \bibinfo {author} {\bibfnamefont {P.}~\bibnamefont {{Hoppe}}},
  \bibinfo {author} {\bibfnamefont {E.}~\bibnamefont {{Zinner}}}, \ and\
  \bibinfo {author} {\bibfnamefont {R.~S.}\ \bibnamefont {{Lewis}}},\
  }\href@noop {} {\bibfield  {journal} {\bibinfo  {journal} {Meteoritics}\
  }\textbf {\bibinfo {volume} {30}},\ \bibinfo {pages} {679} (\bibinfo {year}
  {1995})}\BibitemShut {NoStop}%
\bibitem [{\citenamefont {{Leitner}}\ and\ \citenamefont
  {{Hoppe}}(2022)}]{Leitner22}%
  \BibitemOpen
  \bibfield  {author} {\bibinfo {author} {\bibfnamefont {J.}~\bibnamefont
  {{Leitner}}}\ and\ \bibinfo {author} {\bibfnamefont {P.}~\bibnamefont
  {{Hoppe}}},\ }in\ \href@noop {} {\emph {\bibinfo {booktitle} {LPI
  Contributions}}},\ \bibinfo {series} {LPI Contributions}, Vol.\ \bibinfo
  {volume} {2695}\ (\bibinfo {year} {2022})\ p.\ \bibinfo {pages}
  {6252}\BibitemShut {NoStop}%
\bibitem [{\citenamefont {{Lugaro}}\ \emph {et~al.}(2020)\citenamefont
  {{Lugaro}}, \citenamefont {{Cseh}}, \citenamefont {{Vil{\'a}gos}},
  \citenamefont {{Karakas}}, \citenamefont {{Ventura}}, \citenamefont
  {{Dell'Agli}}, \citenamefont {{Trappitsch}}, \citenamefont {{Hampel}},
  \citenamefont {{D'Orazi}}, \citenamefont {{Pereira}}, \citenamefont
  {{Tagliente}}, \citenamefont {{Szab{\'o}}}, \citenamefont {{Pignatari}},
  \citenamefont {{Battino}}, \citenamefont {{Tattersall}}, \citenamefont
  {{Ek}}, \citenamefont {{Sch{\"o}nb{\"a}chler}}, \citenamefont {{Hron}},\ and\
  \citenamefont {{Nittler}}}]{Lugaro20}%
  \BibitemOpen
  \bibfield  {author} {\bibinfo {author} {\bibfnamefont {M.}~\bibnamefont
  {{Lugaro}}}, \bibinfo {author} {\bibfnamefont {B.}~\bibnamefont {{Cseh}}},
  \bibinfo {author} {\bibfnamefont {B.}~\bibnamefont {{Vil{\'a}gos}}}, \bibinfo
  {author} {\bibfnamefont {A.~I.}\ \bibnamefont {{Karakas}}}, \bibinfo {author}
  {\bibfnamefont {P.}~\bibnamefont {{Ventura}}}, \bibinfo {author}
  {\bibfnamefont {F.}~\bibnamefont {{Dell'Agli}}}, \bibinfo {author}
  {\bibfnamefont {R.}~\bibnamefont {{Trappitsch}}}, \bibinfo {author}
  {\bibfnamefont {M.}~\bibnamefont {{Hampel}}}, \bibinfo {author}
  {\bibfnamefont {V.}~\bibnamefont {{D'Orazi}}}, \bibinfo {author}
  {\bibfnamefont {C.~B.}\ \bibnamefont {{Pereira}}}, \bibinfo {author}
  {\bibfnamefont {G.}~\bibnamefont {{Tagliente}}}, \bibinfo {author}
  {\bibfnamefont {G.~M.}\ \bibnamefont {{Szab{\'o}}}}, \bibinfo {author}
  {\bibfnamefont {M.}~\bibnamefont {{Pignatari}}}, \bibinfo {author}
  {\bibfnamefont {U.}~\bibnamefont {{Battino}}}, \bibinfo {author}
  {\bibfnamefont {A.}~\bibnamefont {{Tattersall}}}, \bibinfo {author}
  {\bibfnamefont {M.}~\bibnamefont {{Ek}}}, \bibinfo {author} {\bibfnamefont
  {M.}~\bibnamefont {{Sch{\"o}nb{\"a}chler}}}, \bibinfo {author} {\bibfnamefont
  {J.}~\bibnamefont {{Hron}}}, \ and\ \bibinfo {author} {\bibfnamefont {L.~R.}\
  \bibnamefont {{Nittler}}},\ }\href@noop {} {\bibfield  {journal} {\bibinfo
  {journal} {The Astrophysical Journal}\ }\textbf {\bibinfo {volume} {898}},\
  \bibinfo {pages} {96} (\bibinfo {year} {2020})}\BibitemShut {NoStop}%
\bibitem [{\citenamefont {K\"appeler}\ \emph {et~al.}(1996)\citenamefont
  {K\"appeler}, \citenamefont {Toukan}, \citenamefont {Schumann},\ and\
  \citenamefont {Mengoni}}]{KAP96}%
  \BibitemOpen
  \bibfield  {author} {\bibinfo {author} {\bibfnamefont {F.}~\bibnamefont
  {K\"appeler}}, \bibinfo {author} {\bibfnamefont {K.~A.}\ \bibnamefont
  {Toukan}}, \bibinfo {author} {\bibfnamefont {M.}~\bibnamefont {Schumann}}, \
  and\ \bibinfo {author} {\bibfnamefont {A.}~\bibnamefont {Mengoni}},\ }\href
  {\doibase https://doi.org/10.1103/PhysRevC.53.1397} {\bibfield  {journal}
  {\bibinfo  {journal} {Phys. Rev. C}\ }\textbf {\bibinfo {volume} {53}},\
  \bibinfo {pages} {1397} (\bibinfo {year} {1996})}\BibitemShut {NoStop}%
\bibitem [{\citenamefont {Harnood}\ \emph {et~al.}(2000)\citenamefont
  {Harnood}, \citenamefont {Igashira}, \citenamefont {Matsumoto}, \citenamefont
  {Mizuno},\ and\ \citenamefont {Ohsaki}}]{HAR00}%
  \BibitemOpen
  \bibfield  {author} {\bibinfo {author} {\bibfnamefont {S.}~\bibnamefont
  {Harnood}}, \bibinfo {author} {\bibfnamefont {M.}~\bibnamefont {Igashira}},
  \bibinfo {author} {\bibfnamefont {T.}~\bibnamefont {Matsumoto}}, \bibinfo
  {author} {\bibfnamefont {S.}~\bibnamefont {Mizuno}}, \ and\ \bibinfo {author}
  {\bibfnamefont {T.}~\bibnamefont {Ohsaki}},\ }\href {\doibase
  10.1080/18811248.2000.9714952} {\bibfield  {journal} {\bibinfo  {journal}
  {Journal of Nuclear Science and Technology}\ }\textbf {\bibinfo {volume}
  {37}},\ \bibinfo {pages} {740} (\bibinfo {year} {2000})},\ \Eprint
  {http://arxiv.org/abs/https://doi.org/10.1080/18811248.2000.9714952}
  {https://doi.org/10.1080/18811248.2000.9714952} \BibitemShut {NoStop}%
\bibitem [{\citenamefont {Amaducci}\ \emph {et~al.}(2021)\citenamefont
  {Amaducci}, \citenamefont {Colonna}, \citenamefont {Cosentino}, \citenamefont
  {Cristallo},\ and\ \citenamefont {et~al.}}]{AMA21}%
  \BibitemOpen
  \bibfield  {author} {\bibinfo {author} {\bibfnamefont {S.}~\bibnamefont
  {Amaducci}}, \bibinfo {author} {\bibfnamefont {N.}~\bibnamefont {Colonna}},
  \bibinfo {author} {\bibfnamefont {L.}~\bibnamefont {Cosentino}}, \bibinfo
  {author} {\bibfnamefont {S.}~\bibnamefont {Cristallo}}, \ and\ \bibinfo
  {author} {\bibnamefont {et~al.}},\ }\href {\doibase 10.3390/universe7060200}
  {\bibfield  {journal} {\bibinfo  {journal} {Universe}\ }\textbf {\bibinfo
  {volume} {7}} (\bibinfo {year} {2021}),\ 10.3390/universe7060200}\BibitemShut
  {NoStop}%
\bibitem [{kad()}]{kadonis}%
  \BibitemOpen
  \href {https://exp-astro.de/kadonis1.0/} {\enquote {\bibinfo {title}
  {Karlsruhe astrophysical database of nucleosynthesis in stars (kadonis
  v1.0)},}\ }\BibitemShut {NoStop}%
\bibitem [{\citenamefont {{Cosner K. and Truran J. W.}}(1981)}]{COS81}%
  \BibitemOpen
  \bibfield  {author} {\bibinfo {author} {\bibnamefont {{Cosner K. and Truran
  J. W.}}},\ }\href@noop {} {\bibfield  {journal} {\bibinfo  {journal}
  {Astrophysics and Space Science}\ }\textbf {\bibinfo {volume} {78}},\
  \bibinfo {pages} {85} (\bibinfo {year} {1981})}\BibitemShut {NoStop}%
\bibitem [{\citenamefont {Takahashi}\ and\ \citenamefont
  {Yokoi}(1987)}]{TAK87}%
  \BibitemOpen
  \bibfield  {author} {\bibinfo {author} {\bibfnamefont {K.}~\bibnamefont
  {Takahashi}}\ and\ \bibinfo {author} {\bibfnamefont {K.}~\bibnamefont
  {Yokoi}},\ }\href {\doibase https://doi.org/10.1016/0092-640X(87)90010-6}
  {\bibfield  {journal} {\bibinfo  {journal} {At. Data Nucl. Data Tables}\
  }\textbf {\bibinfo {volume} {36}},\ \bibinfo {pages} {375 } (\bibinfo {year}
  {1987})}\BibitemShut {NoStop}%
\bibitem [{\citenamefont {Koloczek}\ \emph {et~al.}(2016)\citenamefont
  {Koloczek}, \citenamefont {Thomas}, \citenamefont {Glorius}, \citenamefont
  {Plag}, \citenamefont {Pignatari}, \citenamefont {Reifarth}, \citenamefont
  {Ritter}, \citenamefont {Schmidt},\ and\ \citenamefont
  {Sonnabend}}]{Koloczek16}%
  \BibitemOpen
  \bibfield  {author} {\bibinfo {author} {\bibfnamefont {A.}~\bibnamefont
  {Koloczek}}, \bibinfo {author} {\bibfnamefont {B.}~\bibnamefont {Thomas}},
  \bibinfo {author} {\bibfnamefont {J.}~\bibnamefont {Glorius}}, \bibinfo
  {author} {\bibfnamefont {R.}~\bibnamefont {Plag}}, \bibinfo {author}
  {\bibfnamefont {M.}~\bibnamefont {Pignatari}}, \bibinfo {author}
  {\bibfnamefont {R.}~\bibnamefont {Reifarth}}, \bibinfo {author}
  {\bibfnamefont {C.}~\bibnamefont {Ritter}}, \bibinfo {author} {\bibfnamefont
  {S.}~\bibnamefont {Schmidt}}, \ and\ \bibinfo {author} {\bibfnamefont
  {K.}~\bibnamefont {Sonnabend}},\ }\href@noop {} {\bibfield  {journal}
  {\bibinfo  {journal} {Atomic Data and Nuclear Data Tables}\ }\textbf
  {\bibinfo {volume} {108}},\ \bibinfo {pages} {1} (\bibinfo {year}
  {2016})}\BibitemShut {NoStop}%
\bibitem [{\citenamefont {Young}\ \emph {et~al.}(2003)\citenamefont {Young},
  \citenamefont {Knierman}, \citenamefont {Rigby},\ and\ \citenamefont
  {Arnett}}]{YOU03}%
  \BibitemOpen
  \bibfield  {author} {\bibinfo {author} {\bibfnamefont {P.~A.}\ \bibnamefont
  {Young}}, \bibinfo {author} {\bibfnamefont {K.~A.}\ \bibnamefont {Knierman}},
  \bibinfo {author} {\bibfnamefont {J.~R.}\ \bibnamefont {Rigby}}, \ and\
  \bibinfo {author} {\bibfnamefont {D.}~\bibnamefont {Arnett}},\ }\href
  {\doibase 10.1086/377428} {\bibfield  {journal} {\bibinfo  {journal} {The
  Astrophysical Journal}\ }\textbf {\bibinfo {volume} {595}},\ \bibinfo {pages}
  {1114} (\bibinfo {year} {2003})}\BibitemShut {NoStop}%
\bibitem [{\citenamefont {Roederer}\ \emph {et~al.}(2011)\citenamefont
  {Roederer}, \citenamefont {Marino},\ and\ \citenamefont {Sneden}}]{ROE11}%
  \BibitemOpen
  \bibfield  {author} {\bibinfo {author} {\bibfnamefont {I.~U.}\ \bibnamefont
  {Roederer}}, \bibinfo {author} {\bibfnamefont {A.~F.}\ \bibnamefont
  {Marino}}, \ and\ \bibinfo {author} {\bibfnamefont {C.}~\bibnamefont
  {Sneden}},\ }\href {\doibase 10.1088/0004-637X/742/1/37} {\bibfield
  {journal} {\bibinfo  {journal} {The Astrophysical Journal}\ }\textbf
  {\bibinfo {volume} {742}},\ \bibinfo {pages} {37} (\bibinfo {year}
  {2011})}\BibitemShut {NoStop}%
\bibitem [{\citenamefont {Straniero}\ \emph {et~al.}(2014)\citenamefont
  {Straniero}, \citenamefont {Cristallo},\ and\ \citenamefont
  {Piersanti}}]{STR14}%
  \BibitemOpen
  \bibfield  {author} {\bibinfo {author} {\bibfnamefont {O.}~\bibnamefont
  {Straniero}}, \bibinfo {author} {\bibfnamefont {S.}~\bibnamefont
  {Cristallo}}, \ and\ \bibinfo {author} {\bibfnamefont {L.}~\bibnamefont
  {Piersanti}},\ }\href {\doibase 10.1088/0004-637X/785/1/77} {\bibfield
  {journal} {\bibinfo  {journal} {The Astrophysical Journal}\ }\textbf
  {\bibinfo {volume} {785}},\ \bibinfo {pages} {77} (\bibinfo {year}
  {2014})}\BibitemShut {NoStop}%
\bibitem [{\citenamefont {{Sahoo, Rudra N.}}\ \emph {et~al.}(2023)\citenamefont
  {{Sahoo, Rudra N.}}, \citenamefont {{Tessler, Moshe}}, \citenamefont
  {{Halfon, Shlomi}}, \citenamefont {{Kijel, Dani}}, \citenamefont {{Kreisel,
  Arik}}, \citenamefont {{Paul, Michael}}, \citenamefont {{Shor, Asher}},\ and\
  \citenamefont {{Weissman, Leo}}}]{SAH23}%
  \BibitemOpen
  \bibfield  {author} {\bibinfo {author} {\bibnamefont {{Sahoo, Rudra N.}}},
  \bibinfo {author} {\bibnamefont {{Tessler, Moshe}}}, \bibinfo {author}
  {\bibnamefont {{Halfon, Shlomi}}}, \bibinfo {author} {\bibnamefont {{Kijel,
  Dani}}}, \bibinfo {author} {\bibnamefont {{Kreisel, Arik}}}, \bibinfo
  {author} {\bibnamefont {{Paul, Michael}}}, \bibinfo {author} {\bibnamefont
  {{Shor, Asher}}}, \ and\ \bibinfo {author} {\bibnamefont {{Weissman, Leo}}},\
  }\href {\doibase 10.1051/epjconf/202327906002} {\bibfield  {journal}
  {\bibinfo  {journal} {EPJ Web Conf.}\ }\textbf {\bibinfo {volume} {279}},\
  \bibinfo {pages} {06002} (\bibinfo {year} {2023})}\BibitemShut {NoStop}%
\bibitem [{\citenamefont {Tessler}\ \emph {et~al.}(2016)\citenamefont
  {Tessler}, \citenamefont {Paul}, \citenamefont {Palchan}, \citenamefont
  {Halfon}, \citenamefont {Weissman}, \citenamefont {Hazenshprung},
  \citenamefont {Kreisel}, \citenamefont {Makmal}, \citenamefont {Shor},
  \citenamefont {Silverman}, \citenamefont {Kashiv},\ and\ \citenamefont
  {Purtschert}}]{TES16}%
  \BibitemOpen
  \bibfield  {author} {\bibinfo {author} {\bibfnamefont {M.}~\bibnamefont
  {Tessler}}, \bibinfo {author} {\bibfnamefont {M.}~\bibnamefont {Paul}},
  \bibinfo {author} {\bibfnamefont {T.}~\bibnamefont {Palchan}}, \bibinfo
  {author} {\bibfnamefont {S.}~\bibnamefont {Halfon}}, \bibinfo {author}
  {\bibfnamefont {L.}~\bibnamefont {Weissman}}, \bibinfo {author}
  {\bibfnamefont {N.}~\bibnamefont {Hazenshprung}}, \bibinfo {author}
  {\bibfnamefont {A.}~\bibnamefont {Kreisel}}, \bibinfo {author} {\bibfnamefont
  {T.}~\bibnamefont {Makmal}}, \bibinfo {author} {\bibfnamefont
  {A.}~\bibnamefont {Shor}}, \bibinfo {author} {\bibfnamefont {I.}~\bibnamefont
  {Silverman}}, \bibinfo {author} {\bibfnamefont {Y.}~\bibnamefont {Kashiv}}, \
  and\ \bibinfo {author} {\bibfnamefont {R.}~\bibnamefont {Purtschert}},\ }in\
  \href@noop {} {\emph {\bibinfo {booktitle} {Proceedings of the 26th Int.
  Nuclear Physics Conf.}}}\ (\bibinfo {address} {Adelaide, Australia},\
  \bibinfo {year} {2016})\ p.\ \bibinfo {pages} {139},\ \bibinfo {note} {url =
  {https://pos.sissa.it/281/139/pdf}}\BibitemShut {NoStop}%
\bibitem [{\citenamefont {Mardor}\ \emph {et~al.}(2018)\citenamefont {Mardor},
  \citenamefont {Aviv}, \citenamefont {Avrigeanu}, \citenamefont {Berkovits},
  \citenamefont {Dahan}, \citenamefont {Dickel}, \citenamefont {Eliyahu},
  \citenamefont {Gai}, \citenamefont {Gavish-Segev}, \citenamefont {Halfon},
  \citenamefont {Hass}, \citenamefont {Hirsh}, \citenamefont {Kaiser},
  \citenamefont {Kijel}, \citenamefont {Kreisel}, \citenamefont {Mishnayot},
  \citenamefont {Mukul}, \citenamefont {Ohayon}, \citenamefont {Paul},
  \citenamefont {Perry}, \citenamefont {Rahangdale}, \citenamefont {Rodnizki},
  \citenamefont {Ron}, \citenamefont {Sasson-Zukran}, \citenamefont {Shor},
  \citenamefont {Silverman}, \citenamefont {Tessler}, \citenamefont
  {Vaintraub},\ and\ \citenamefont {Weissman}}]{MAR18}%
  \BibitemOpen
  \bibfield  {author} {\bibinfo {author} {\bibfnamefont {I.}~\bibnamefont
  {Mardor}}, \bibinfo {author} {\bibfnamefont {O.}~\bibnamefont {Aviv}},
  \bibinfo {author} {\bibfnamefont {M.}~\bibnamefont {Avrigeanu}}, \bibinfo
  {author} {\bibfnamefont {D.}~\bibnamefont {Berkovits}}, \bibinfo {author}
  {\bibfnamefont {A.}~\bibnamefont {Dahan}}, \bibinfo {author} {\bibfnamefont
  {T.}~\bibnamefont {Dickel}}, \bibinfo {author} {\bibfnamefont
  {I.}~\bibnamefont {Eliyahu}}, \bibinfo {author} {\bibfnamefont
  {M.}~\bibnamefont {Gai}}, \bibinfo {author} {\bibfnamefont {I.}~\bibnamefont
  {Gavish-Segev}}, \bibinfo {author} {\bibfnamefont {S.}~\bibnamefont
  {Halfon}}, \bibinfo {author} {\bibfnamefont {M.}~\bibnamefont {Hass}},
  \bibinfo {author} {\bibfnamefont {T.}~\bibnamefont {Hirsh}}, \bibinfo
  {author} {\bibfnamefont {B.}~\bibnamefont {Kaiser}}, \bibinfo {author}
  {\bibfnamefont {D.}~\bibnamefont {Kijel}}, \bibinfo {author} {\bibfnamefont
  {A.}~\bibnamefont {Kreisel}}, \bibinfo {author} {\bibfnamefont
  {Y.}~\bibnamefont {Mishnayot}}, \bibinfo {author} {\bibfnamefont
  {I.}~\bibnamefont {Mukul}}, \bibinfo {author} {\bibfnamefont
  {B.}~\bibnamefont {Ohayon}}, \bibinfo {author} {\bibfnamefont
  {M.}~\bibnamefont {Paul}}, \bibinfo {author} {\bibfnamefont {A.}~\bibnamefont
  {Perry}}, \bibinfo {author} {\bibfnamefont {H.}~\bibnamefont {Rahangdale}},
  \bibinfo {author} {\bibfnamefont {J.}~\bibnamefont {Rodnizki}}, \bibinfo
  {author} {\bibfnamefont {G.}~\bibnamefont {Ron}}, \bibinfo {author}
  {\bibfnamefont {R.}~\bibnamefont {Sasson-Zukran}}, \bibinfo {author}
  {\bibfnamefont {A.}~\bibnamefont {Shor}}, \bibinfo {author} {\bibfnamefont
  {I.}~\bibnamefont {Silverman}}, \bibinfo {author} {\bibfnamefont
  {M.}~\bibnamefont {Tessler}}, \bibinfo {author} {\bibfnamefont
  {S.}~\bibnamefont {Vaintraub}}, \ and\ \bibinfo {author} {\bibfnamefont
  {L.}~\bibnamefont {Weissman}},\ }\href {\doibase 10.1140/epja/i2018-12526-2}
  {\bibfield  {journal} {\bibinfo  {journal} {Eur. Phys. J. A}\ }\textbf
  {\bibinfo {volume} {54}},\ \bibinfo {pages} {91} (\bibinfo {year}
  {2018})}\BibitemShut {NoStop}%
\bibitem [{\citenamefont {Paul}\ \emph {et~al.}(2019)\citenamefont {Paul},
  \citenamefont {Tessler}, \citenamefont {Friedman}, \citenamefont {Halfon},
  \citenamefont {Palchan}, \citenamefont {Weissman}, \citenamefont {Arenshtam},
  \citenamefont {Berkovits}, \citenamefont {Eisen}, \citenamefont {Eliahu},
  \citenamefont {Feinberg}, \citenamefont {Kijel}, \citenamefont {Kreisel},
  \citenamefont {Mardor}, \citenamefont {Shimel}, \citenamefont {Shor},\ and\
  \citenamefont {Silverman}}]{PAU19a}%
  \BibitemOpen
  \bibfield  {author} {\bibinfo {author} {\bibfnamefont {M.}~\bibnamefont
  {Paul}}, \bibinfo {author} {\bibfnamefont {M.}~\bibnamefont {Tessler}},
  \bibinfo {author} {\bibfnamefont {M.}~\bibnamefont {Friedman}}, \bibinfo
  {author} {\bibfnamefont {S.}~\bibnamefont {Halfon}}, \bibinfo {author}
  {\bibfnamefont {T.}~\bibnamefont {Palchan}}, \bibinfo {author} {\bibfnamefont
  {L.}~\bibnamefont {Weissman}}, \bibinfo {author} {\bibfnamefont
  {A.}~\bibnamefont {Arenshtam}}, \bibinfo {author} {\bibfnamefont
  {D.}~\bibnamefont {Berkovits}}, \bibinfo {author} {\bibfnamefont
  {Y.}~\bibnamefont {Eisen}}, \bibinfo {author} {\bibfnamefont
  {I.}~\bibnamefont {Eliahu}}, \bibinfo {author} {\bibfnamefont
  {G.}~\bibnamefont {Feinberg}}, \bibinfo {author} {\bibfnamefont
  {D.}~\bibnamefont {Kijel}}, \bibinfo {author} {\bibfnamefont
  {A.}~\bibnamefont {Kreisel}}, \bibinfo {author} {\bibfnamefont
  {I.}~\bibnamefont {Mardor}}, \bibinfo {author} {\bibfnamefont
  {G.}~\bibnamefont {Shimel}}, \bibinfo {author} {\bibfnamefont
  {A.}~\bibnamefont {Shor}}, \ and\ \bibinfo {author} {\bibfnamefont
  {I.}~\bibnamefont {Silverman}},\ }\href {\doibase
  https://doi.org/10.1140/epja/i2019-12723-5} {\bibfield  {journal} {\bibinfo
  {journal} {Eur. Phys. J. A}\ }\textbf {\bibinfo {volume} {55}},\ \bibinfo
  {pages} {44} (\bibinfo {year} {2019})}\BibitemShut {NoStop}%
\bibitem [{\citenamefont {Ratynski}\ and\ \citenamefont
  {K\"{a}ppeler}(1988)}]{RAT88}%
  \BibitemOpen
  \bibfield  {author} {\bibinfo {author} {\bibfnamefont {W.}~\bibnamefont
  {Ratynski}}\ and\ \bibinfo {author} {\bibfnamefont {F.}~\bibnamefont
  {K\"{a}ppeler}},\ }\href {\doibase https://doi.org/10.1103/PhysRevC.37.595}
  {\bibfield  {journal} {\bibinfo  {journal} {Phys. Rev. C}\ }\textbf {\bibinfo
  {volume} {37}},\ \bibinfo {pages} {595} (\bibinfo {year} {1988})}\BibitemShut
  {NoStop}%
\bibitem [{\citenamefont {Friedman}\ \emph {et~al.}(2013)\citenamefont
  {Friedman}, \citenamefont {Cohen}, \citenamefont {Paul}, \citenamefont
  {Berkovits}, \citenamefont {Eisen}, \citenamefont {Feinberg}, \citenamefont
  {Giorginis}, \citenamefont {Halfon}, \citenamefont {Krása}, \citenamefont
  {Plompen},\ and\ \citenamefont {Shor}}]{FRI13}%
  \BibitemOpen
  \bibfield  {author} {\bibinfo {author} {\bibfnamefont {M.}~\bibnamefont
  {Friedman}}, \bibinfo {author} {\bibfnamefont {D.}~\bibnamefont {Cohen}},
  \bibinfo {author} {\bibfnamefont {M.}~\bibnamefont {Paul}}, \bibinfo {author}
  {\bibfnamefont {D.}~\bibnamefont {Berkovits}}, \bibinfo {author}
  {\bibfnamefont {Y.}~\bibnamefont {Eisen}}, \bibinfo {author} {\bibfnamefont
  {G.}~\bibnamefont {Feinberg}}, \bibinfo {author} {\bibfnamefont
  {G.}~\bibnamefont {Giorginis}}, \bibinfo {author} {\bibfnamefont
  {S.}~\bibnamefont {Halfon}}, \bibinfo {author} {\bibfnamefont
  {A.}~\bibnamefont {Krása}}, \bibinfo {author} {\bibfnamefont
  {A.}~\bibnamefont {Plompen}}, \ and\ \bibinfo {author} {\bibfnamefont
  {A.}~\bibnamefont {Shor}},\ }\href {\doibase
  http://dx.doi.org/10.1016/j.nima.2012.09.027} {\bibfield  {journal} {\bibinfo
   {journal} {Nucl. Instrum. Methods Phys. Res. A}\ }\textbf {\bibinfo {volume}
  {698}},\ \bibinfo {pages} {117 } (\bibinfo {year} {2013})}\BibitemShut
  {NoStop}%
\bibitem [{\citenamefont {Agostinelli}\ \emph {et~al.}(2003)\citenamefont
  {Agostinelli} \emph {et~al.}}]{AGO03}%
  \BibitemOpen
  \bibfield  {author} {\bibinfo {author} {\bibfnamefont {S.}~\bibnamefont
  {Agostinelli}} \emph {et~al.},\ }\href {\doibase
  http://dx.doi.org/10.1016/S0168-9002(03)01368-8} {\bibfield  {journal}
  {\bibinfo  {journal} {Nucl. Instrum. Methods Phys. Res. Sect. A}\ }\textbf
  {\bibinfo {volume} {506}},\ \bibinfo {pages} {250 } (\bibinfo {year}
  {2003})}\BibitemShut {NoStop}%
\bibitem [{\citenamefont {Tessler}\ \emph {et~al.}(2015)\citenamefont
  {Tessler}, \citenamefont {Paul}, \citenamefont {Arenshtam}, \citenamefont
  {Feinberg}, \citenamefont {Friedman}, \citenamefont {Halfon}, \citenamefont
  {Kijel}, \citenamefont {Weissman}, \citenamefont {Aviv}, \citenamefont
  {Berkovits}, \citenamefont {Eisen}, \citenamefont {Eliyahu}, \citenamefont
  {Haquin}, \citenamefont {Kreisel}, \citenamefont {Mardor}, \citenamefont
  {Shimel}, \citenamefont {Shor}, \citenamefont {Silverman},\ and\
  \citenamefont {Yungrais}}]{TES15}%
  \BibitemOpen
  \bibfield  {author} {\bibinfo {author} {\bibfnamefont {M.}~\bibnamefont
  {Tessler}}, \bibinfo {author} {\bibfnamefont {M.}~\bibnamefont {Paul}},
  \bibinfo {author} {\bibfnamefont {A.}~\bibnamefont {Arenshtam}}, \bibinfo
  {author} {\bibfnamefont {G.}~\bibnamefont {Feinberg}}, \bibinfo {author}
  {\bibfnamefont {M.}~\bibnamefont {Friedman}}, \bibinfo {author}
  {\bibfnamefont {S.}~\bibnamefont {Halfon}}, \bibinfo {author} {\bibfnamefont
  {D.}~\bibnamefont {Kijel}}, \bibinfo {author} {\bibfnamefont
  {L.}~\bibnamefont {Weissman}}, \bibinfo {author} {\bibfnamefont
  {O.}~\bibnamefont {Aviv}}, \bibinfo {author} {\bibfnamefont {D.}~\bibnamefont
  {Berkovits}}, \bibinfo {author} {\bibfnamefont {Y.}~\bibnamefont {Eisen}},
  \bibinfo {author} {\bibfnamefont {I.}~\bibnamefont {Eliyahu}}, \bibinfo
  {author} {\bibfnamefont {G.}~\bibnamefont {Haquin}}, \bibinfo {author}
  {\bibfnamefont {A.}~\bibnamefont {Kreisel}}, \bibinfo {author} {\bibfnamefont
  {I.}~\bibnamefont {Mardor}}, \bibinfo {author} {\bibfnamefont
  {G.}~\bibnamefont {Shimel}}, \bibinfo {author} {\bibfnamefont
  {A.}~\bibnamefont {Shor}}, \bibinfo {author} {\bibfnamefont {I.}~\bibnamefont
  {Silverman}}, \ and\ \bibinfo {author} {\bibfnamefont {Z.}~\bibnamefont
  {Yungrais}},\ }\href {\doibase
  https://doi.org/10.1016/j.physletb.2015.10.058} {\bibfield  {journal}
  {\bibinfo  {journal} {Phys. Lett. B}\ }\textbf {\bibinfo {volume} {751}},\
  \bibinfo {pages} {418 } (\bibinfo {year} {2015})}\BibitemShut {NoStop}%
\bibitem [{NND()}]{NNDC}%
  \BibitemOpen
  \href {https://www.nndc.bnl.gov/nudat3/} {\bibinfo  {journal}
  {https://www.nndc.bnl.gov/nudat3/}\ }\BibitemShut {NoStop}%
\bibitem [{\citenamefont {Zahnow}\ \emph {et~al.}(1995)\citenamefont {Zahnow},
  \citenamefont {Angulo}, \citenamefont {Rolfs}, \citenamefont {Schmidt},
  \citenamefont {Schulte},\ and\ \citenamefont {Somorjai}}]{ZAH95}%
  \BibitemOpen
\bibfield  {journal} {  }\bibfield  {author} {\bibinfo {author} {\bibfnamefont
  {D.}~\bibnamefont {Zahnow}}, \bibinfo {author} {\bibfnamefont
  {C.}~\bibnamefont {Angulo}}, \bibinfo {author} {\bibfnamefont
  {C.}~\bibnamefont {Rolfs}}, \bibinfo {author} {\bibfnamefont
  {S.}~\bibnamefont {Schmidt}}, \bibinfo {author} {\bibfnamefont {W.~H.}\
  \bibnamefont {Schulte}}, \ and\ \bibinfo {author} {\bibfnamefont
  {E.}~\bibnamefont {Somorjai}},\ }\href {\doibase
  https://doi.org/10.1007/BF01289534} {\bibfield  {journal} {\bibinfo
  {journal} {Zeitschrift f{\"u}r Physik A Hadrons and Nuclei}\ }\textbf
  {\bibinfo {volume} {351}},\ \bibinfo {pages} {229} (\bibinfo {year}
  {1995})}\BibitemShut {NoStop}%
\bibitem [{\citenamefont {Munch}\ \emph {et~al.}(2018)\citenamefont {Munch},
  \citenamefont {{Sølund Kirsebom}}, \citenamefont {Swartz}, \citenamefont
  {Riisager},\ and\ \citenamefont {Fynbo}}]{MUN18}%
  \BibitemOpen
  \bibfield  {author} {\bibinfo {author} {\bibfnamefont {M.}~\bibnamefont
  {Munch}}, \bibinfo {author} {\bibfnamefont {O.}~\bibnamefont {{Sølund
  Kirsebom}}}, \bibinfo {author} {\bibfnamefont {J.~A.}\ \bibnamefont
  {Swartz}}, \bibinfo {author} {\bibfnamefont {K.}~\bibnamefont {Riisager}}, \
  and\ \bibinfo {author} {\bibfnamefont {H.~O.~U.}\ \bibnamefont {Fynbo}},\
  }\href {\doibase https://doi.org/10.1016/j.physletb.2018.06.013} {\bibfield
  {journal} {\bibinfo  {journal} {Physics Letters B}\ }\textbf {\bibinfo
  {volume} {782}},\ \bibinfo {pages} {779} (\bibinfo {year}
  {2018})}\BibitemShut {NoStop}%
\bibitem [{\citenamefont {Brown}\ \emph {et~al.}(2018)\citenamefont {Brown},
  \citenamefont {Chadwick}, \citenamefont {Capote}, \citenamefont {Kahler},
  \citenamefont {Trkov}, \citenamefont {Herman}, \citenamefont {Sonzogni},
  \citenamefont {Danon}, \citenamefont {Carlson}, \citenamefont {Dunn},
  \citenamefont {Smith}, \citenamefont {Hale}, \citenamefont {Arbanas},
  \citenamefont {Arcilla}, \citenamefont {Bates}, \citenamefont {Beck},
  \citenamefont {Becker}, \citenamefont {Brown}, \citenamefont {Casperson},
  \citenamefont {Conlin}, \citenamefont {Cullen}, \citenamefont {Descalle},
  \citenamefont {Firestone}, \citenamefont {Gaines}, \citenamefont {Guber},
  \citenamefont {Hawari}, \citenamefont {Holmes}, \citenamefont {Johnson},
  \citenamefont {Kawano}, \citenamefont {Kiedrowski}, \citenamefont {Koning},
  \citenamefont {Kopecky}, \citenamefont {Leal}, \citenamefont {Lestone},
  \citenamefont {Lubitz}, \citenamefont {Damián}, \citenamefont {Mattoon},
  \citenamefont {McCutchan}, \citenamefont {Mughabghab}, \citenamefont
  {Navratil}, \citenamefont {Neudecker}, \citenamefont {Nobre}, \citenamefont
  {Noguere}, \citenamefont {Paris}, \citenamefont {Pigni}, \citenamefont
  {Plompen}, \citenamefont {Pritychenko}, \citenamefont {Pronyaev},
  \citenamefont {Roubtsov}, \citenamefont {Rochman}, \citenamefont {Romano},
  \citenamefont {Schillebeeckx}, \citenamefont {Simakov}, \citenamefont {Sin},
  \citenamefont {Sirakov}, \citenamefont {Sleaford}, \citenamefont {Sobes},
  \citenamefont {Soukhovitskii}, \citenamefont {Stetcu}, \citenamefont {Talou},
  \citenamefont {Thompson}, \citenamefont {van~der Marck}, \citenamefont
  {Welser-Sherrill}, \citenamefont {Wiarda}, \citenamefont {White},
  \citenamefont {Wormald}, \citenamefont {Wright}, \citenamefont {Zerkle},
  \citenamefont {Zerovnik},\ and\ \citenamefont {Zhu}}]{BRO18}%
  \BibitemOpen
  \bibfield  {author} {\bibinfo {author} {\bibfnamefont {D.}~\bibnamefont
  {Brown}}, \bibinfo {author} {\bibfnamefont {M.}~\bibnamefont {Chadwick}},
  \bibinfo {author} {\bibfnamefont {R.}~\bibnamefont {Capote}}, \bibinfo
  {author} {\bibfnamefont {A.}~\bibnamefont {Kahler}}, \bibinfo {author}
  {\bibfnamefont {A.}~\bibnamefont {Trkov}}, \bibinfo {author} {\bibfnamefont
  {M.}~\bibnamefont {Herman}}, \bibinfo {author} {\bibfnamefont
  {A.}~\bibnamefont {Sonzogni}}, \bibinfo {author} {\bibfnamefont
  {Y.}~\bibnamefont {Danon}}, \bibinfo {author} {\bibfnamefont
  {A.}~\bibnamefont {Carlson}}, \bibinfo {author} {\bibfnamefont
  {M.}~\bibnamefont {Dunn}}, \bibinfo {author} {\bibfnamefont {D.}~\bibnamefont
  {Smith}}, \bibinfo {author} {\bibfnamefont {G.}~\bibnamefont {Hale}},
  \bibinfo {author} {\bibfnamefont {G.}~\bibnamefont {Arbanas}}, \bibinfo
  {author} {\bibfnamefont {R.}~\bibnamefont {Arcilla}}, \bibinfo {author}
  {\bibfnamefont {C.}~\bibnamefont {Bates}}, \bibinfo {author} {\bibfnamefont
  {B.}~\bibnamefont {Beck}}, \bibinfo {author} {\bibfnamefont {B.}~\bibnamefont
  {Becker}}, \bibinfo {author} {\bibfnamefont {F.}~\bibnamefont {Brown}},
  \bibinfo {author} {\bibfnamefont {R.}~\bibnamefont {Casperson}}, \bibinfo
  {author} {\bibfnamefont {J.}~\bibnamefont {Conlin}}, \bibinfo {author}
  {\bibfnamefont {D.}~\bibnamefont {Cullen}}, \bibinfo {author} {\bibfnamefont
  {M.-A.}\ \bibnamefont {Descalle}}, \bibinfo {author} {\bibfnamefont
  {R.}~\bibnamefont {Firestone}}, \bibinfo {author} {\bibfnamefont
  {T.}~\bibnamefont {Gaines}}, \bibinfo {author} {\bibfnamefont
  {K.}~\bibnamefont {Guber}}, \bibinfo {author} {\bibfnamefont
  {A.}~\bibnamefont {Hawari}}, \bibinfo {author} {\bibfnamefont
  {J.}~\bibnamefont {Holmes}}, \bibinfo {author} {\bibfnamefont
  {T.}~\bibnamefont {Johnson}}, \bibinfo {author} {\bibfnamefont
  {T.}~\bibnamefont {Kawano}}, \bibinfo {author} {\bibfnamefont
  {B.}~\bibnamefont {Kiedrowski}}, \bibinfo {author} {\bibfnamefont
  {A.}~\bibnamefont {Koning}}, \bibinfo {author} {\bibfnamefont
  {S.}~\bibnamefont {Kopecky}}, \bibinfo {author} {\bibfnamefont
  {L.}~\bibnamefont {Leal}}, \bibinfo {author} {\bibfnamefont {J.}~\bibnamefont
  {Lestone}}, \bibinfo {author} {\bibfnamefont {C.}~\bibnamefont {Lubitz}},
  \bibinfo {author} {\bibfnamefont {J.~M.}\ \bibnamefont {Damián}}, \bibinfo
  {author} {\bibfnamefont {C.}~\bibnamefont {Mattoon}}, \bibinfo {author}
  {\bibfnamefont {E.}~\bibnamefont {McCutchan}}, \bibinfo {author}
  {\bibfnamefont {S.}~\bibnamefont {Mughabghab}}, \bibinfo {author}
  {\bibfnamefont {P.}~\bibnamefont {Navratil}}, \bibinfo {author}
  {\bibfnamefont {D.}~\bibnamefont {Neudecker}}, \bibinfo {author}
  {\bibfnamefont {G.}~\bibnamefont {Nobre}}, \bibinfo {author} {\bibfnamefont
  {G.}~\bibnamefont {Noguere}}, \bibinfo {author} {\bibfnamefont
  {M.}~\bibnamefont {Paris}}, \bibinfo {author} {\bibfnamefont
  {M.}~\bibnamefont {Pigni}}, \bibinfo {author} {\bibfnamefont
  {A.}~\bibnamefont {Plompen}}, \bibinfo {author} {\bibfnamefont
  {B.}~\bibnamefont {Pritychenko}}, \bibinfo {author} {\bibfnamefont
  {V.}~\bibnamefont {Pronyaev}}, \bibinfo {author} {\bibfnamefont
  {D.}~\bibnamefont {Roubtsov}}, \bibinfo {author} {\bibfnamefont
  {D.}~\bibnamefont {Rochman}}, \bibinfo {author} {\bibfnamefont
  {P.}~\bibnamefont {Romano}}, \bibinfo {author} {\bibfnamefont
  {P.}~\bibnamefont {Schillebeeckx}}, \bibinfo {author} {\bibfnamefont
  {S.}~\bibnamefont {Simakov}}, \bibinfo {author} {\bibfnamefont
  {M.}~\bibnamefont {Sin}}, \bibinfo {author} {\bibfnamefont {I.}~\bibnamefont
  {Sirakov}}, \bibinfo {author} {\bibfnamefont {B.}~\bibnamefont {Sleaford}},
  \bibinfo {author} {\bibfnamefont {V.}~\bibnamefont {Sobes}}, \bibinfo
  {author} {\bibfnamefont {E.}~\bibnamefont {Soukhovitskii}}, \bibinfo {author}
  {\bibfnamefont {I.}~\bibnamefont {Stetcu}}, \bibinfo {author} {\bibfnamefont
  {P.}~\bibnamefont {Talou}}, \bibinfo {author} {\bibfnamefont
  {I.}~\bibnamefont {Thompson}}, \bibinfo {author} {\bibfnamefont
  {S.}~\bibnamefont {van~der Marck}}, \bibinfo {author} {\bibfnamefont
  {L.}~\bibnamefont {Welser-Sherrill}}, \bibinfo {author} {\bibfnamefont
  {D.}~\bibnamefont {Wiarda}}, \bibinfo {author} {\bibfnamefont
  {M.}~\bibnamefont {White}}, \bibinfo {author} {\bibfnamefont
  {J.}~\bibnamefont {Wormald}}, \bibinfo {author} {\bibfnamefont
  {R.}~\bibnamefont {Wright}}, \bibinfo {author} {\bibfnamefont
  {M.}~\bibnamefont {Zerkle}}, \bibinfo {author} {\bibfnamefont
  {G.}~\bibnamefont {Zerovnik}}, \ and\ \bibinfo {author} {\bibfnamefont
  {Y.}~\bibnamefont {Zhu}},\ }\href {\doibase
  https://doi.org/10.1016/j.nds.2018.02.001} {\bibfield  {journal} {\bibinfo
  {journal} {Nucl. Data Sheets}\ }\textbf {\bibinfo {volume} {148}},\ \bibinfo
  {pages} {1 } (\bibinfo {year} {2018})}\BibitemShut {NoStop}%
\bibitem [{\citenamefont {Lederer}\ \emph {et~al.}(2011)\citenamefont
  {Lederer}, \citenamefont {Colonna}, \citenamefont {Domingo-Pardo},
  \citenamefont {Gunsing}, \citenamefont {K\"{a}ppeler} \emph
  {et~al.}}]{LED11shrt}%
  \BibitemOpen
  \bibfield  {author} {\bibinfo {author} {\bibfnamefont {C.}~\bibnamefont
  {Lederer}}, \bibinfo {author} {\bibfnamefont {N.}~\bibnamefont {Colonna}},
  \bibinfo {author} {\bibfnamefont {C.}~\bibnamefont {Domingo-Pardo}}, \bibinfo
  {author} {\bibfnamefont {F.}~\bibnamefont {Gunsing}}, \bibinfo {author}
  {\bibnamefont {K\"{a}ppeler}},  \emph {et~al.},\ }\href {\doibase
  10.1103/PhysRevC.83.034608} {\bibfield  {journal} {\bibinfo  {journal} {Phys.
  Rev. C}\ }\textbf {\bibinfo {volume} {83}},\ \bibinfo {pages} {034608}
  (\bibinfo {year} {2011})}\BibitemShut {NoStop}%
\bibitem [{\citenamefont {Massimi}\ \emph {et~al.}(2014)\citenamefont
  {Massimi}, \citenamefont {Becker}, \citenamefont {Dupont}, \citenamefont
  {Kopecky}, \citenamefont {Lampoudis}, \citenamefont {Massarczyk},
  \citenamefont {Moxon}, \citenamefont {Pronyaev}, \citenamefont
  {Schillebeeckx}, \citenamefont {Sirakov},\ and\ \citenamefont
  {Wynants}}]{MAS14}%
  \BibitemOpen
  \bibfield  {author} {\bibinfo {author} {\bibfnamefont {C.}~\bibnamefont
  {Massimi}}, \bibinfo {author} {\bibfnamefont {B.}~\bibnamefont {Becker}},
  \bibinfo {author} {\bibfnamefont {E.}~\bibnamefont {Dupont}}, \bibinfo
  {author} {\bibfnamefont {S.}~\bibnamefont {Kopecky}}, \bibinfo {author}
  {\bibfnamefont {C.}~\bibnamefont {Lampoudis}}, \bibinfo {author}
  {\bibfnamefont {R.}~\bibnamefont {Massarczyk}}, \bibinfo {author}
  {\bibfnamefont {M.}~\bibnamefont {Moxon}}, \bibinfo {author} {\bibfnamefont
  {V.}~\bibnamefont {Pronyaev}}, \bibinfo {author} {\bibfnamefont
  {P.}~\bibnamefont {Schillebeeckx}}, \bibinfo {author} {\bibfnamefont
  {I.}~\bibnamefont {Sirakov}}, \ and\ \bibinfo {author} {\bibfnamefont
  {R.}~\bibnamefont {Wynants}},\ }\href {\doibase
  https://doi.org/10.1140/epja/i2014-14124-8} {\bibfield  {journal} {\bibinfo
  {journal} {Eur. Phys. J. A}\ }\textbf {\bibinfo {volume} {50}},\ \bibinfo
  {pages} {124} (\bibinfo {year} {2014})}\BibitemShut {NoStop}%
\bibitem [{\citenamefont {Feinberg}\ \emph {et~al.}(2012)\citenamefont
  {Feinberg}, \citenamefont {Friedman}, \citenamefont {Kr\'{a}sa},
  \citenamefont {Shor}, \citenamefont {Eisen}, \citenamefont {Berkovits},
  \citenamefont {Cohen}, \citenamefont {Giorginis}, \citenamefont {Hirsh},
  \citenamefont {Paul}, \citenamefont {Plompen},\ and\ \citenamefont
  {Tsuk}}]{FEI12}%
  \BibitemOpen
  \bibfield  {author} {\bibinfo {author} {\bibfnamefont {G.}~\bibnamefont
  {Feinberg}}, \bibinfo {author} {\bibfnamefont {M.}~\bibnamefont {Friedman}},
  \bibinfo {author} {\bibfnamefont {A.}~\bibnamefont {Kr\'{a}sa}}, \bibinfo
  {author} {\bibfnamefont {A.}~\bibnamefont {Shor}}, \bibinfo {author}
  {\bibfnamefont {Y.}~\bibnamefont {Eisen}}, \bibinfo {author} {\bibfnamefont
  {D.}~\bibnamefont {Berkovits}}, \bibinfo {author} {\bibfnamefont
  {D.}~\bibnamefont {Cohen}}, \bibinfo {author} {\bibfnamefont
  {G.}~\bibnamefont {Giorginis}}, \bibinfo {author} {\bibfnamefont
  {T.}~\bibnamefont {Hirsh}}, \bibinfo {author} {\bibfnamefont
  {M.}~\bibnamefont {Paul}}, \bibinfo {author} {\bibfnamefont {A.~J.~M.}\
  \bibnamefont {Plompen}}, \ and\ \bibinfo {author} {\bibfnamefont
  {E.}~\bibnamefont {Tsuk}},\ }\href {\doibase
  https://doi.org/10.1103/PhysRevC.85.055810} {\bibfield  {journal} {\bibinfo
  {journal} {Phys. Rev. C}\ }\textbf {\bibinfo {volume} {85}},\ \bibinfo
  {pages} {055810} (\bibinfo {year} {2012})}\BibitemShut {NoStop}%
\end{thebibliography}
%merlin.mbs apsrev4-1.bst 2010-07-25 4.21a (PWD, AO, DPC) hacked
%Control: key (0)
%Control: author (72) initials jnrlst
%Control: editor formatted (1) identically to author
%Control: production of article title (-1) disabled
%Control: page (0) single
%Control: year (1) truncated
%Control: production of eprint (0) enabled
%

\end{document}